\documentclass[10pt,draftcls,onecolumn]{IEEEtran}
\IEEEoverridecommandlockouts                              

\usepackage{soul}
\usepackage{graphics} 
\usepackage[pdftex]{graphicx}
\usepackage{epsfig} 
\usepackage{mathptmx} 
\usepackage{times} 
\usepackage{amsmath} 
\usepackage{amssymb}  
\usepackage{subfigure}
\usepackage{verbatim}
\usepackage{cite}
\usepackage{epstopdf}
\usepackage{color}
\usepackage{relsize}
\usepackage{url}
\usepackage{wasysym}
\usepackage{nopageno}
\definecolor{brown}{rgb}{0.59, 0.29, 0.0}
\definecolor{purple}{rgb}{0.63, 0.36, 0.94}
\newtheorem{theorem}{Theorem}

\newtheorem{assumption}{Assumption}
\usepackage{color}

\newtheorem{corollary}{Corollary}

\newtheorem{lemma}{Lemma}

\newtheorem{proposition}{Proposition}

{ 
\newtheorem{remark}{Remark}

}

\newcommand*{\vpointer}{\vcenter{\hbox{\scalebox{1}{\Huge\pointer}}}}
\newcounter{prob1}
\newcounter{prob2}
\newcounter{prob3}
\newcounter{prob4}
\newcounter{prob5}
\newcounter{prob6}
\newcounter{prob7}

\title{\LARGE \bf Youla Coding and Computation of Gaussian Feedback Capacity}

\author{\quad Chong Li and Nicola Elia %
\thanks{Dr. ~Chong Li is with Qualcomm Research, Bridgewater, NJ, 08807.
chongl@qti.qualcomm.com}
\thanks{Dr.~Nicola Elia is with the Department of Electrical and
Computer Engineering,
Iowa State University,
Ames, IA, 50011 nelia@iastate.edu}
\thanks{This paper was presented (in part) in \cite{chong2015}.} %
}

\begin{document}

\maketitle \thispagestyle{empty} \pagestyle{plain}
\begin{abstract}
In this paper, we propose an approach to numerically compute the feedback capacity of stationary finite dimensional Gaussian channels and construct (arbitrarily close to) capacity-achieving feedback codes.
In particular, we first extend the interpretation of feedback communication over stationary finite dimensional Gaussian channels as feedback control systems by showing that,
the problem of finding stabilizing feedback controllers with maximal reliable transmission rate over Youla parameters coincides with the problem of finding strictly causal filters to achieve feedback capacity derived in \cite{Kim10}.
This extended interpretation provides an approach to construct deterministic feedback coding schemes with double exponential decaying error probability. We next propose asymptotic capacity-achieving upper bounds, which can be numerically evaluated by solving finite dimensional convex optimizations.
From the filters that achieve the upper bounds, we apply the Youla-based interpretation to construct feasible filters, i.e., feedback codes, leading to a sequence of lower bounds.
We prove the sequence of lower bounds is asymptotically capacity-achieving.
\end{abstract}


\setcounter{prob1}{1}
\setcounter{prob2}{2}
\setcounter{prob3}{3}
\setcounter{prob4}{4}
\setcounter{prob5}{5}
\setcounter{prob6}{6}
\setcounter{prob7}{7}

\section{Introduction}
To facilitate the reading, we first introduce some notations as follows.

{\bf Notation:} Uppercase and corresponding lowercase letters $(e.g. Y,Z,y,z)$ denote random variables and realizations, respectively.
$\log$ denotes the logarithm base $2$ and $0\log0=0$. $C^\infty_{[a,b]}$ refers to the set of bounded continuous functions on $[a,b]$.
 We use $\mathbf{x}'$ to denote the transpose of a  real vector or matrix $\mathbf{x}$.

$\mathcal{RH}_2={\mathcal{RH}}_\infty$  denotes the set of real-rational transfer functions corresponding to the $z-$transform of the impulse response of  linear time invariant finite-dimensional (LTI-FD) causal stable systems \cite{Bruce_note}
\footnote{In general,  $\mathcal{H}_{\infty} \subset \mathcal{H}_2$, for discrete-time systems. When restricted  to the real-rational functions, however, $\mathcal{RH}_{\infty} = \mathcal{RH}_2$. Note that \cite{Bruce_note} uses $\lambda$-transform instead of the $z-$tranform, where $\lambda=1/z$.}.
We interchangeably use $\mathcal{RH}_{\infty}$ and $\mathcal{RH}_2$ throughout the paper. A function $f(z)\in \mathcal{RH}_2$ has all its poles strictly inside the unit disc.

In this paper, we consider a discrete-time Gaussian channel with noiseless feedback.
\begin{assumption}\label{ass0.ass}
The additive Gaussian channel is modeled as
\begin{equation}\label{model: forward channel}
Y_i=U_i+W_i, \qquad i=1,2,\cdots
\end{equation}
where the Gaussian noise sequence $\lbrace W_i \rbrace_{i=1}^{\infty}$ is assumed to be wide-sense stationary and has power spectral density $\mathbb{S}_{w}(e^{j\theta})>0$ for $\forall \theta\in [-\pi,\pi)$. Moreover, {the power spectral density satisfies the \textit{Paley-Wiener} condition,}
$$\frac{1}{2\pi}\int_{-\pi}^{\pi}|\log \mathbb{S}_{w}(e^{j\theta})|d\theta< \infty.$$
\end{assumption}

Unless the contrary is explicitly stated, we use the term ``stationary'' to refer to stationary in wide sense.

For a code with achievable rate $R$, we specify a sequence of $(n,2^{nR})$ channel codes as follows. $M$ is a uniformly distributed message index where $M=m\in\lbrace 1,2,3,\cdots,2^{nR}\rbrace$. There exists an encoding process $U_i(M,Y^{i-1})$, where $Y^{i-1}=\lbrace Y_0, Y_1,\cdots, Y_{i-1}\rbrace$, for $i = 1,2,\cdots,n$ and $U_1(M, Y^0) = U_1(M)$ with average transmit power constraint. That is, the channel input $U_i$ is determined by the message index $M$ and the previous channel output $Y^{i-1}$. Furthermore, there exists a decoding function $g$: $Y^n\rightarrow \lbrace 1,2,\cdots,2^{nR}\rbrace$ with an error probability satisfying
$P_e^{(n)}=\frac{1}{2^{nR}}\sum_{m=1}^{2^{nR}} P(g(Y^n)\neq m | M=m)\leq \epsilon_n$,
where $\lim_{n\rightarrow\infty}\epsilon_n=0$. The objective of communication is to deliver $M$ to the receiver at highest code rate with arbitrarily small error probability. The feedback capacity $C_{fb}$ is defined as the supremium of all achievable rates $R$.
\indent 

As shown in \cite{Kim10}, the feedback capacity of channel (\ref{model: forward channel}) with the average power budget $P$ can be characterized by
\begin{equation}
\begin{split}
C_{fb}=&\max_{\mathbb{S}_V,\mathbb{Q}}\frac{1}{2\pi}\int_{-\pi}^{\pi}\frac{1}{2}\log\left( \frac{\mathbb{S}_v(e^{j\theta})}{\mathbb{S}_w(e^{j\theta})}+|1+\mathbb{Q}(e^{j\theta})|^2\right)d\theta,\\
s.t. \quad  &\frac{1}{2\pi}\int_{-\pi}^{\pi}\mathbb{S}_v(e^{j\theta})+|\mathbb{Q}(e^{j\theta})|^2\mathbb{S}_w(e^{j\theta})d\theta\leq P,\\
& \mathbb{S}_v(e^{j\theta})\geq 0,\\
& \mathbb{Q}(e^{j\theta}) = \sum_{k=1}^{\infty} q_k e^{jk\theta} \in \mathcal{H}_2\; \text{is strictly causal}.\\
\end{split}
\label{capacity_general}
\end{equation}

\begin{assumption}\label{LTIFD.ass}
In this paper, we further assume that noise $W$ is the output of a LTI-FD stable system $\mathbb{H} \in \mathcal{RH}_2$, not necessarily minimum phase,  driven by white Gaussian noise with zero mean and unit variance.
The power spectral density (PSD) of $W$  has a canonical spectral factorization given by  ${\mathbb S}_w(e^{j\theta})= |\mathbb{H}(e^{j\theta})|^2$.
\end{assumption}

Note that any stationary process can be approximated with arbitrary accuracy by this LTI-FD filtering model and this approximation is very ``efficient'' as it corresponds to the rational approximation of the spectral density \cite{book.stat.linear.sys}.

Given the capacity of white Gaussian channel with feedback has been well known, in this paper we consider non-white channel noise spectra.

\begin{assumption}\label{nonwhite.ass}
Noise $W$ has a colored (non-white) power spectral density.
\end{assumption}
\begin{remark}\label{sv0.rem} Under Assumption \ref{LTIFD.ass} and \ref{nonwhite.ass}, \cite{Kim10} showed that the optimal solution to (\ref{capacity_general}) must have $\mathbb{S}_v=0$. Furthermore, Proposition 5.1 in \cite{Kim10} presented that the capacity is necessarily achieved by a \textit{rational} filter $\mathbb{Q}$. Therefore, without loss of optimality, we restrict the search space $\mathbb{Q}$ to $\mathcal{RH}_2$ in (\ref{capacity_general}).
\end{remark}

Specifically, (\ref{capacity_general}) can be simplified to
\begin{equation}
\begin{split}
C_{fb}=&\max_{\mathbb{Q}}\frac{1}{2\pi}\int_{-\pi}^{\pi}\log |1+\mathbb{Q}(e^{j\theta})|d\theta,\\
s.t. \quad  &\frac{1}{2\pi}\int_{-\pi}^{\pi}|\mathbb{Q}(e^{j\theta})|^2\mathbb{S}_w(e^{j\theta})d\theta\leq P,\\
& \mathbb{Q}(e^{j\theta}) = \sum_{k=1}^{\infty} q_k e^{jk\theta} \in \mathcal{RH}_2\; \text{is strictly causal} .\\
\end{split}
\label{capacity_short01}
\end{equation}
\begin{remark} \label{rem:unit_zero}
Under Assumption  \ref{LTIFD.ass} and \ref{nonwhite.ass}, the optimal $\mathbb{Q}$ has no zeros on the unit circle (Proposition 5.1 (ii)  in \cite{Kim10}).
\end{remark}

While the above characterizations are elegant, they are infinite dimensional. As stated in \cite{Kim10},\\

\noindent\textit{``$\cdots$ except for the first-order ARMA spectrum, it is still a nontrivial task to find analytically (or even numerically) the optimal filter and corresponding feedback capacity''.}\\

\indent In this paper, we aim to solve the above optimization problem and explicitly construct capacity-achieving feedback codes. Firstly, we revisit and extend the interpretation of  feedback communication over Gaussian channels as feedback control problems \cite{unified.theory}. In particular, we highlight the central role of \textit{Youla parameterization over all stabilizing controllers} in connecting these two theories by showing that  the characterization of the maximum-rate over all stabilizing controllers and the feedback capacity over all coding schemes coincide. This new interpretation provides an approach to construct an explicit (sub-)optimal communication scheme (i.e. encoder and decoder) directly from the filter $\mathbb{Q}$ in (\ref{capacity_short01}). Next, to find an optimal $\mathbb{Q}$, we provide an alternative characterization of the feedback capacity, from which an asymptotic capacity-achieving sequence of upper bounds is derived and can be numerically evaluated by solving finite dimensional convex optimizations. Furthermore, from the filters $\mathbb{Q}$ that achieve upper bounds, we derive a sequence of lower bounds on the feedback capacity by constructing deterministic feedback codes with double exponential decaying error probability. Essentially, this constructed coding scheme has the structure of a generalized Schalkwijk-Kailath scheme which have been extensively studied by Elia \cite{Elia2004}, Kim\cite{Kim10}, Liu-Elia\cite{Liu_CIS}, Shayevitz-Feder\cite{posterior.matching}, Ardestanizadeh-Minero-Franceschetti \cite{Franceschetti_contrl_comm_fd}\cite{massimo_lqg} and others. It is proved that the sequence of lower bounds converges to the capacity, leading to an asymptotically optimal feedback coding scheme. It is worth noting that \cite{Kim10} (see Theorem 6.1 and Lemma 6.1) has shown a structural, not computable result that a $k$-dimensional generalization of the Schalkwijk-Kailath coding scheme achieves the feedback capacity for any auto-regression moving average (ARMA) noise spectrum of order $k$, while we herein provide a numerically explicit feedback coding scheme from $\mathbb{Q}$ by leveraging control-oriented derivations.

\subsection{Related Work}
We review the literature along two avenues: information theory and feedback control theory. As a complete survey is vast and most of them are out of the scope of our discussion, we herein list most relevant results to this paper. In the field of information theory, the investigation on feedback Gaussian capacity has been experiencing a decade journey. \cite{Elias1956} and its sequel \cite{Elias1967} are recognized as the first works on feedback Gaussian channels by proposing feedback coding schemes. \cite{Schalkwijk66} \cite{Schalkwijk66_2} developed an elegant linear feedback coding scheme of achieving the capacity of additive white Gaussian noise (AWGN) channel with noiseless feedback. Thereafter, several works by Butman \cite{Butman69}, \cite{Butman76}, Tiernan\cite{Tiernan74}\cite{Tiernan76}, Wolfowitz\cite{Wolfowitz75} and Ozarow\cite{Ozarow_random90}\cite{Ozarow_upper90} extended this notable result to ARMA Gaussian channels, with an objective to find channel capacity and optimal feedback codes. As a consequence, many interesting upper and lower bounds were obtained. Based on the insight/results from aforementioned literature, \cite{cover89} made a major breakthrough on characterizing the $n$-block capacity of arbitrary feedback Gaussian channels by using asymptotic equipartition property (AEP) theorem. It was also shown that feedback capacity for arbitrary Gaussian channels cannot be increased by factor two or half bit. This $n$-block capacity was extended to the case of feedback Gaussian channels with noisy feedback where capacity bounds and other interesting results were obtained\cite{Love10,chong11_ISIT,Chong11_allerton_upperbound,Chong12_allerton_sideInfo,chong.thesis}. As hinted by this $n$-block capacity characterization, \cite{Kim10} developed a variational characterization on the capacity of stationary feedback Gaussian channels, which is an infinite dimensional optimization problem. For the first-order ARMA noise, this variational characterization yields a closed-form solution on the capacity and shows the optimality of the Schalkwijk-Kailath scheme.

In the field of feedback control theory, many control-based technical tools have been utilized to attack the problem of finding feedback Gaussian channel capacity and capacity-achieving codes. \cite{Elia2004} proposed the derivation of  feedback communication schemes based on a feedback control method.
These results were  obtained from considering the problem of stabilization of a given unstable plant over a Gaussian communication channel.  The communication rate (in the sense of Shannon) over the channel was connected to the degree on instability of the plant. The minimal transmission power for a given unstable plant was obtained by solving the classical ${\cal H}_2$ (or Linear Quadratic Gaussian) problem. However, plants with the same degree of instability may require different transmission power to be stabilized.  \cite{Elia2004} provided the plants that can be stabilized most efficiently, i.e. with the least transmission power for a given degree of instability for special case channels. This approach provides a method of finding feedback coding scheme for Gaussian channels. The approach has been further extended to Markov channels with channel state information\cite{Liu04_ISIT}, connected to the classical linear quadratic Gaussian (LQG) control problem \cite{Franceschetti_contrl_comm_fd}\cite{massimo_lqg}. Specifically, \cite{massimo_lqg} presented a code for an $k$-receiver additive white Gaussian noise broadcast channel with feedback and characterized its sum-rate performance by using the tools from LQG control theory.  Finally, \cite{Liu_CIS} extended the convergence of the fundamental limitations of control and communication to include the limitations of estimation. In light of this unified framework, a set of achievable rates of feedback Gaussian channels were obtained by constructing specific feedback coding schemes via control-oriented approaches.
\cite{Yang_feedbackCapacity} converted the problem of finding feedback Gaussian channel capacity into a form of stochastic control and used dynamic programming to compute the $n$-block capacity.
%


\section{Preliminaries}
In this section we formalize some preliminary results, which directly follow from \cite{Kim10} and are useful for our derivations. In addition, we briefly review the theory of Youla parameterization, a useful tool to construct capacity-achieving feedback codes.
\subsection{Capacity Characterization Revisited}

From (\ref{capacity_general}), it is immediate that
\begin{equation}
\begin{split}
C_{fb}=&\sup_{\mathbb{S}_V,\mathbb{Q}}\frac{1}{2\pi}\int_{-\pi}^{\pi}\frac{1}{2}\log\left( \frac{\mathbb{S}_v(e^{j\theta})}{\mathbb{S}_w(e^{j\theta})}+|1+\mathbb{Q}(e^{j\theta})|^2\right)d\theta,\\
s.t. \quad  &\frac{1}{2\pi}\int_{-\pi}^{\pi}\mathbb{S}_v(e^{j\theta})+|\mathbb{Q}(e^{j\theta})|^2\mathbb{S}_w(e^{j\theta})d\theta\leq P,\\
& \mathbb{S}_v(e^{j\theta})\geq 0,\\
& \mathbb{Q} \in \mathcal{RH}_2\; \text{is strictly causal},\\
\end{split}
\label{capacity_long01}
\end{equation}
since ${\cal RH}_2$ is a dense subspace of ${\cal H}_2$.

\begin{proposition} Under Assumption \ref{LTIFD.ass} the ``sup'' in (\ref{capacity_long01}) is achieved.
\end{proposition}
\begin{IEEEproof}
If the noise spectrum is non-white, then the result follows from (\ref{capacity_short01}) and $\mathbb{S}_v$ must be equal to $0$. If the noise spectrum is white, it is well known that feedback does not change the capacity.  Thus, we can let $\mathbb{S}_v$ be given by the classical water-filling solution and $\mathbb{Q}=0$\footnote{Note that there may exist infinite number of solutions with $\mathbb{Q}\neq 0 \in \mathcal{RH}_2$ in this case.}
\end{IEEEproof}

{ One will see in Section \ref{sec.interpretation} that the above capacity characterization with $\mathbb{Q} \in \mathcal{RH}_2$ (equivalently, $\mathcal{RH}_{\infty}$) is useful to connect feedback communications and feedback control, a crucial step to derive the capacity-achieving feedback codes.}

\subsection{Youla Parameterization}
\begin{figure}
\begin{center}
\includegraphics[scale=0.35]{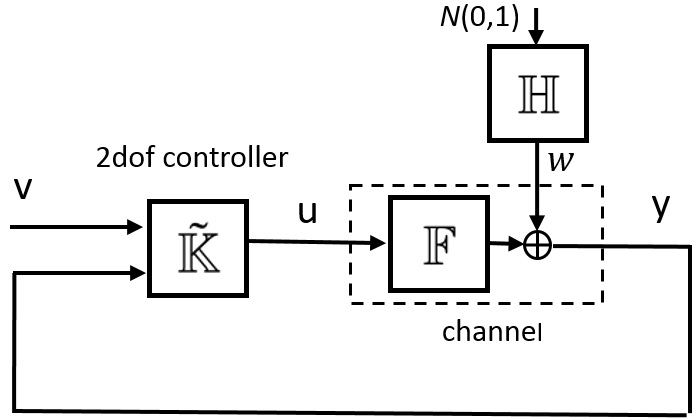}
\caption{Feedback Gaussian channels with two degree of freedom controllers and exogenous input. }
\label{noisyfb_scheme}
\end{center}
\end{figure}

We consider the closed loop set-up shown in Figure \ref{noisyfb_scheme} where ${\mathbb{F}}$ is a single-input-single-output (SISO) LTI-FD plant represented by a rational transfer function, the additive disturbance $w$ is generated according to Assumption \ref{LTIFD.ass}, $\tilde{\mathbb K}$ is a two-degree of freedom (2dof) stabilizing
controller, and $v$ is an exogenous input with power spectral density ${\mathbb S}_v$.
In what follows, we review a fundamental result of linear control theory: Youla-Ku\u{c}era parameterization (also simply known as \textit{Youla parameterization} or \textit{Q-parameterization}) \cite{Doyle92}.
 In particular, we consider the set of two degree of freedom stabilizing controllers for a given plant, and the resulting achievable closed loop maps.
\begin{lemma} (\textit{Youla Parameterization of Controllers}, { Chapter 5.2 in \cite{Doyle92}})
Given a SISO plant $\mathbb{F}$, there exist ${\mathbb{N,D,X,U}}\in {\mathcal{RH}}_\infty$, such  that
$$ {\mathbb{F}}= \frac{\mathbb{N}}{\mathbb{D}},$$
with
\begin{equation}{\label{lemma:youla}}
\mathbb{N}\mathbb{U}+\mathbb{D}\mathbb{X} = 1.
\end{equation}

The above factorization (\ref{lemma:youla}) of plant $\mathbb{F}$ is called coprime factorization. Moreover, the set of all 2dof rational controllers $\tilde{\mathbb{K}}:\,\left[\begin{array}{c} v \cr y\end{array}\right]\to u$,
that stabilize the closed loop of plant $\mathbb{F}$ is given by { (Chapter 5.4, Theorem 2 in \cite{Doyle92})\footnote{Note that the controller parameterizaiton in \cite{Doyle92} is given by $\mathbb{X}-\mathbb{N}\mathbb{Q}$ with a negative sign on $\mathbb{N}\mathbb{Q}$. This should not be confusing as the closed-loop systems considered in \cite{Doyle92} has negative feedback while it is \textit{positive} in our model.}}
$$
\tilde{\mathbb{K}}= \frac{1}{\mathbb{X}+\mathbb{N}\mathbb{Q}}\left[\mathbb{Q}_v,\,(\mathbb{U}+\mathbb{D}\mathbb{Q})\right],\,\mbox{for }  \mathbb{Q}_v\in \mathcal{RH}_\infty, \mathbb{Q} \in \mathcal{RH}_\infty.
$$
\end{lemma}
Note that the order of  the controller is  not restricted in any way.
The main benefit of Youla paramterization will be seen later in the next section. All omitted technical proofs in the following sections are provided in Appendix.

\section{Feedback Control Interpretation of Feedback Capacity for Gaussian Channels }\label{sec.interpretation}

In this section, we propose a control-theoretic approach to derive the feedback capacity formula for the finite dimensional LTI Gaussian channels. The proposed approach based on \cite{Elia2004} reveals the essential role of Youla-parameter in connecting the theories of feedback communications and feedback control, and provides feasible feedback communication schemes with guaranteed transmission rate. As will be proved in the paper, this constructed coding scheme achieves (arbitrarily close to) the feedback capacity.




Consider a channel
\begin{equation}
Y_i=U_i+W_i,
\label{channel.LTI}
\end{equation}
where $W_i$ satisfies Assumption \ref{LTIFD.ass}. As shown in Fig. \ref{noisyfb_scheme}, we are interested in the closed loop stabilization problem over the given channel.
Following {\cite{Elia2004}, we consider the map from $U_i$ to $Y_i$ as the plant, $\mathbb{F}$, where $\mathbb{F}=1$ is stable.

Now, according to Youla parametrization of stabilizing controllers, let $\mathbb{D}=\mathbb{X}=1$, $\mathbb{N}=1$ and $\mathbb{U}=0$. Then all the 2dof LTI-FD stabilizing controllers for the plant $\mathbb{F}=1$ have the following expression,
represented as transfer functions:
\begin{equation}\label{youla.eq}
\tilde{\mathbb{K}}=\left[\mathbb{K}_v,\,\mathbb{K}\right]=(1+\mathbb{Q})^{-1}\left[\mathbb{Q}_v,\, \mathbb{Q}\right]
\end{equation}
where $\mathbb{Q}\in \mathcal{RH}_\infty, \mathbb{Q}_v \in \mathcal{RH}_\infty$.

Working with ${\mathbb{Q}}$ (and $\mathbb{Q}_v$) instead of $\tilde{\mathbb K}$ is more convenient.
The main advantage comes from the fact that the above transformation convexifies the set of achievable closed loop maps by a stabilizing controller.
In particular,
\begin{equation}\label{Y.eq}
\begin{array}{ll}
Y&=W+\mathbb{K}_v V+\mathbb{K}Y\\
&=(1-\mathbb{K})^{-1}(W+\mathbb{K}_v V)\\
&\stackrel{(a)}{=}(1+\mathbb{Q})W+\mathbb{Q}_vV,
\end{array}
\end{equation}
where step $(a)$ follows from (\ref{youla.eq}). Similarly,
\begin{equation}\label{U.eq}
U=Y-W=\mathbb{Q}W+\mathbb{Q}_vV.
\end{equation}
\begin{remark}
In what follows, we restrict our attention to strictly causal feedback operations. This is equivalent to restricting ${\mathbb Q}$ to be strictly proper. From (\ref{U.eq}), we can see that this way $u$ is not instantaneous function of $w$.
\end{remark}


In information theory of feedback communication systems, it is known that directed information \footnote{Directed information, first defined by Massey \cite{Massey1990}, has been widely used in characterizing the capacity of channels with feedback \cite{Yang_finteState,kim08,Tati09,chong_isit11_capacity,Kramer_thesis}. Moreover, it has interpretation on portfolio theory, data compression and hypothesis testing \cite{Permuter11}.}  from channel inputs to outputs measures the causal information transfer for channels with noiseless feedback and can be used to characterize the feedback channel capacity. In our model, the directed information from channel inputs $U^T$ to channel outputs $Y^T$ is defined by
$$
I(U^T\rightarrow Y^T)=\sum_{i=1}^{T} I(U^i;Y_i|Y^{i-1}),
$$
where $I(U^i;Y_i|Y^{i-1})$ denotes the conditional mutual information. We next characterize the average directed information and the average power of channel input $U$ in terms of the Youla parameters.
\begin{lemma}\label{lem.direct.info.power} Given the Youla parametrization in (\ref{youla.eq}) with $\mathbb{Q}$ strictly proper,
\begin{equation*}
\begin{split}
\displaystyle\lim_{T\to \infty}\frac{1}{T}I(U^T\rightarrow Y^T)=&\frac{1}{4\pi} \displaystyle\int_{-\pi}^\pi\log\left(\frac{|\mathbb{Q}_v(e^{j\theta})|^2\mathbb{S}_v(e^{j\theta})}{\mathbb{S}_w(e^{j\theta})}+|1+\mathbb{Q}(e^{j\theta})|^2\right)d\theta,\\
\lim_{T\rightarrow \infty} \frac{1}{T}\sum_{t=1}^{T}U_t^2 =&\frac{1}{2\pi}\int_{-\pi}^\pi |\mathbb{Q}(e^{j\theta})|^2{\mathbb S}_w(e^{j\theta})+|\mathbb{Q}_v(e^{j\theta})|^2\mathbb{S}_v(e^{j\theta})d\theta.\\
\end{split}
\end{equation*}
\end{lemma}

{ Note that the derivation of the characterization of the above averaged directed information mainly follows from the proof of Theorem 1 in \cite{Elia2004}. The detailed proof of this lemma can be found in Appendix. With this lemma in hand, we are now interested in finding the largest directed information rate for the closed-loop stabilization under constrained average power, by searching over $\mathbb{Q}\in {\mathcal{RH}}_\infty$ strictly proper, $\mathbb{Q}_v\in {\mathcal{RH}}_\infty$ and PSD $\mathbb{S}_v$.
We see, however, that $\mathbb{Q}_v$ can simply be chosen to be equal to $1$. Putting above together, we can directly obtain the following theorem.}

\begin{theorem}\label{capacity_rec.thm}
Consider an additive Gaussian channel in (\ref{channel.LTI}) under Assumption \ref{LTIFD.ass}. Given the average channel input power budget $P$, the largest directed information rate (in the sense of Shannon) over all strictly causal LTI stabilizing controllers can be characterized by (\ref{capacity_long01}).
\end{theorem}
{ This theorem indicates that the feedback capacity characterization (\ref{capacity_long01}) derived from information theory can be equivalently derived from a control approach based on Youla parameterization. According to  Remark \ref{sv0.rem} and Assumption \ref{nonwhite.ass}, one must have $\mathbb{S}_v(e^{j\theta}) = 0$. As a consequence, we obtain (\ref{capacity_short01}).}

In summary, the above derivation
\begin{enumerate}
\item extends the feedback control interpretation of feedback communication system over Gaussian channels with access to feedback and shows how the Youla parameter $\mathbb{Q}$ is central to the feedback capacity problem;
\item motivates, in the next section, an explicit construction of capacity-achieving feedback coding schemes from Youla parameter ${\mathbb Q}$, resolving an issue left open in \cite{Kim10}.
\end{enumerate}

\begin{figure}
\begin{center}
\includegraphics[scale=0.35]{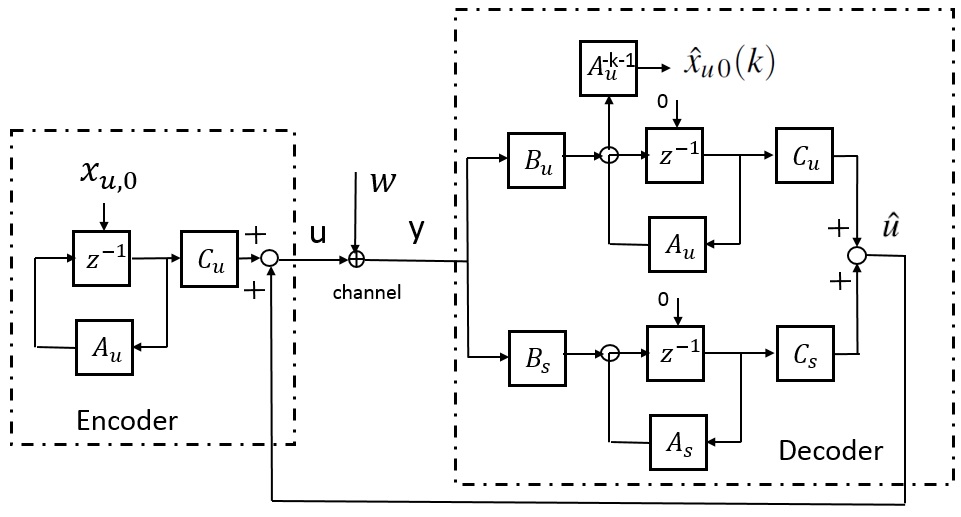}
\caption{Decomposition of controller $\mathbb{K}$ into feedback encoder and decoder.}
\label{fig:codingStructure}
\end{center}
\end{figure}

\subsection{Construction of Capacity-achieving Feedback Codes from Youla Parameter $\mathbb{Q}$} \label{subsection::feasible coding scheme}
Once a feasible $\mathbb{Q}$ is found for the above optimization (\ref{capacity_short01}), which is possibly optimal or arbitrarily close to optimal, we construct a controller
$\mathbb{K}=\mathbb{Q}(1+\mathbb{Q})^{-1}$  as defined in (\ref{youla.eq}) to stabilize the channel within the prescribed input average power limit.

We next show  how to construct a feasible feedback coding scheme from $\mathbb{K}$, a scheme which is deterministic (time-invariant) and has double exponential decaying decoding error probability. We follow \cite{Elia2004}.

$\mathbb{K}$ can always be realized with the following observable and controllable state-space realization (Chapter 3 in \cite{book.modern.control}):
\begin{equation}
\begin{split}
\mathbb{K}: \qquad \begin{bmatrix} X_s(k+1) \\ X_u(k+1)\end{bmatrix} &= \begin{bmatrix} A_s & 0 \\ 0 & A_u\end{bmatrix} \begin{bmatrix} X_s(k) \\ X_u(k)\end{bmatrix} + \begin{bmatrix} B_s \\ B_u\end{bmatrix}Y(k)\\
U(k) &= \begin{bmatrix} C_s & C_u\end{bmatrix} \begin{bmatrix} X_s(k) \\ X_u(k)\end{bmatrix}.\\
\end{split}
\label{codingScheme_SS}
\end{equation}
Based on Remark \ref{rem:unit_zero}, we assume that the eigenvalues of $A_u$ are strictly outside the unit disc while the eigenvalues of $A_s$ are strictly inside the unit disc. Without loss of generality we assume that $A_s$ and $A_u$ are in Jordan form. Assume $A_u$ has $m$ eigenvalues, denoted by $\lambda_i(A_u)$, $i = 1,2,\cdots, m$.

Starting with the decoder, we decompose ${\mathbb K}$ as follows. We present the simplest solution here, others are possible. This coding scheme is illustrated in Fig.\ref{fig:codingStructure}.\\

{\em \bf Linear Decoder}\\
The decoder runs ${\mathbb K}$ driven by $Y$.
$$
\begin{array}{cccl}
X_s(k+1)&=&A_sX_s(k)+B_sY(k),\; &X_{s}(0)=0.\\
\hat{X}_u(k+1)&=&A_u\hat{X}_u(k)+B_uY(k),\; &\hat{X}_{u}(0)=0.
\end{array}
$$
It produces two signals:
an estimate of the negative initial condition of the encoder
$$
\hat{X}_{u\,0}(k)=A_u^{-k-1}\hat{X}_u(k+1).
$$
and a feedback signal
$$
\hat{U}(k) = \begin{bmatrix} C_s & C_u\end{bmatrix} \begin{bmatrix} X_s(k) \\ \hat{X}_u(k)\end{bmatrix}.\\
$$

{\em \bf Linear Encoder}\\
The encoder runs the following dynamics
$$
\begin{array}{rcl}
\tilde{X}_u(k+1)&=&A_u\tilde{X}_u(k),\;\tilde{X}_u(0)=X_{u,0},\\
\tilde{U}_u(k)&=&C_u\tilde{X}_u(k),
\end{array}
$$
where $X_{u,0}$ represents the message index. It receives $\hat{U}$ and produces the channel input
$$
\begin{array}{rcl}
U(k)&=&\tilde{U}_u(k)+\hat{U}(k)\\
&=& C_u (\tilde{X}_u(k)+\hat{X}_u(k)) + C_s X_s(k).
\end{array}
$$
Since the closed loop is stable,  $U(k)$ goes to zero with time if the noise is not present. Given the system is observable, this implies that
$\hat{X}_u(k)\to -\tilde{X}_u(k)$. Thus,
$-\hat{X}_{u\,0}(k)$ is an  estimate at time $k$ of $\tilde{X}_u(0)=X_{u,0}$.

In the presence of noise, the above coding scheme leads to $\hat{X}_{u\,0}(k) \backsim \mathcal{N}(-X_{u,0},A_u^{-k}\mathbb{E}[\hat{X}_u(k)\hat{X}_u(k)'] (A_u^{-k})')$ for large $k$, where $\mathbb{E}[\hat{X}_u(k)\hat{X}_u(k)']$ represents the state covariance matrix. Note that, since the system is observable and controllable, the matrix $\mathbb{E}[\hat{X}_u(k)\hat{X}_u(k)']$ is positive definite and so is $A_u^{-k}\mathbb{E}[\hat{X}_u(k)\hat{X}_u(k)'] (A_u^{-k})'$ (Theorem 4.3 in \cite{Elia2004}).
{
\begin{remark} The above proposed coding scheme can be viewed as a generalized Schalkwijk-Kailath scheme. Based on the Schalkwijk's scheme in \cite{Schalkwijk66_2}, the channel input (encoder) and the message estimate (decoder) for $k\geq 2$ can be summarized as follows by using the notations in this paper.
\begin{equation}\label{SK scheme}
\begin{split}
U(k) =& \sqrt{A_u^2-1}A_u^{k-1}(\hat{X}_{u\,0}(k-1) + X_{u\,0}),\\
\hat{X}_{u\,0}(k) = &\hat{X}_{u\,0}(k-1) - A_u^{-k-1}\sqrt{A_u^2-1}Y(k),\\
\end{split}
\end{equation}
where  $A_u = \sqrt{\frac{P+\sigma_w^2}{\sigma_w^2}}$ and $\sigma_w$ is the variance of the additive white Gaussian noise in the forward channel \footnote{We clarify that in our scheme $\hat{X}_{u\,0}(k)$ is the estimate of $-X_{u\,0}$, or $-\theta$ in \cite{Schalkwijk66_2}, at time instance $k$.}. Now, if in our scheme we let $$A_u = \sqrt{\frac{P+\sigma_w^2}{\sigma_w^2}}, \quad B_u = -\frac{\sqrt{A_u^2-1}}{A_u},\quad C_u = \sqrt{A_u^2-1},$$ and $A_s=B_s = C_s = 0$, then the channel input and the message estimate in our scheme become identical to the Schalkwijk's scheme. Specifically, based on the proposed linear coding scheme, we have
\begin{equation}\label{equ:input_u_orthoganal}
\begin{split}
U(k)=&\tilde{U}_u(k)+\hat{U}(k)\\
=& C_u(\tilde{X}_u(k)+\hat{X}_u(k))\\
=& C_u (A_u^{k} X_{u\,0}+  A_u^{k} \hat{X}_{u\,0}(k-1))\\
=& C_u A_u^{k} (X_{u\,0}+ \hat{X}_{u\,0}(k-1))\\
=&\sqrt{A_u^2-1}A_u^{k}(\hat{X}_{u\,0}(k-1) + X_{u\,0}).\\
\end{split}
\end{equation}
\begin{equation}\label{equ:hat_msg}
\begin{split}
\hat{X}_{u\,0}(k)=&A_u^{-k-1}\hat{X}_u(k+1)\\
=& A_u^{-k-1}(A_u\hat{X}_u(k)+B_u Y(k)) \\
=& A_u^{-k}\hat{X}_u(k)+A_u^{-k-1}B_u Y(k) \\
=& \hat{X}_{u\,0}(k-1)+A_u^{-k-1}B_u Y(k) \\
= &\hat{X}_{u\,0}(k-1) - A_u^{-k-2}\sqrt{A_u^2-1}Y(k).\\
\end{split}
\end{equation}
By scaling the message $X_{u\,0}$ and the corresponding estimate $\hat{X}_{u\,0}$ by factor $A_u$, we recover the dynamics of the Schalkwijk's scheme. Note that this constant scaling on the message index $X_{u\,0}$ have no effect on the reliable transmission rate and the power cost at channel input. Furthermore, the first transmission instance ($k=1$) of Schalkwijk's scheme in \cite{Schalkwijk66_2} is differentiated from the above dynamics such that the scheme is optimal for all transmission instances. Our scheme, however, is optimal in the steady state which is determined by the above dynamics.
\end{remark}
}
The next theorem describes how fast messages associated with each $X_{u,0}$ are transferred to $-\hat{X}_{u\,0}(k)$
in the presence of the channel noise.

\begin{theorem}
Consider an additive Gaussian channel in (\ref{channel.LTI}) with Assumption \ref{LTIFD.ass} and  \ref{nonwhite.ass}. Given a strictly causal stable Youla parameter $\mathbb{Q}(e^{j\theta})\in \mathcal{RH}_\infty$, the coding scheme described above based on the decomposition of $\mathbb{K} = \mathbb{Q}(1+\mathbb{Q})^{-1}$ achieves a reliable transmission rate (in the sense of Shannon) at
\begin{equation*}
\frac{1}{2\pi}\int_{-\pi}^{\pi}\log |1+\mathbb{Q}(e^{j\theta})|d\theta =\sum_{i=1}^{m} \log|\lambda_i(A_u)| \quad  \textit{bits/channel use}
\end{equation*}
and has double exponential decaying error probability.
\label{thm_capacity_achieving_code}
\end{theorem}

The proof is omitted as it directly follows Theorem 4.3 in \cite{Elia2004}. Notice that the above achievable rate has the same form as the objective function in (\ref{capacity_short01}), implying that a capacity-achieving feedback code can be constructed from $\mathbb{K}$ if $\mathbb{Q}$ is an optimal solution.

In summary, the above discussion provides an approach to construct a feasible feedback coding scheme with rate $\frac{1}{2\pi}\int_{-\pi}^{\pi}\log |1+\mathbb{Q}(e^{j\theta})|d\theta $ over a stationary finite dimensional Gaussian channel, by leveraging Youla parameter $\mathbb{Q}$. However, we need first to obtain an optimal $\mathbb{Q}$ by solving optimization (\ref{capacity_short01}), which is an infinite dimensional non-convex optimization problem.
In the next section, we provide an approach, by solving finite dimensional convex problems, to find an asymptotic capacity-achieving upper bounds on the capacity.
From the filter $\mathbb{Q}$ to achieve the upper bounds, we can construct asymptotically capacity-achieving feedback codes as described in this section.

\section{Upper Bounds on Capacity}
In this section, we first present an alternative characterization of Gaussian feedback capacity by leveraging the inverse Fourier transform. Based on this characterization, a sequence of asymptotic capacity-achieving upper bounds is proposed and evaluated by solving finite dimensional convex optimization problems.
\subsection{Alternative characterization of Gaussian Feedback Capacity}
\indent We focus on the optimization problem (\ref{capacity_short01}). In what follows, we characterize the Gaussian feedback capacity by imposing the causality constraints in terms of the inverse Fourier transform. 
\begin{lemma}\label{lemma_symmetric_filter}
Under Assumption \ref{LTIFD.ass} and \ref{nonwhite.ass},
let $\mathbb{Q}(e^{j\theta}) = a(\theta)+jb(\theta)$, the feedback capacity can be characterized by
\begin{equation}
\begin{split}
C_{fb}=&\max_{\Gamma}\frac{1}{4\pi}\int_{-\pi}^{\pi}\log ((1+a(\theta))^2+ b(\theta)^2 )d\theta\\
s.t. \quad &\frac{1}{2\pi}\int_{-\pi}^{\pi}\left(a^2(\theta)+b^2(\theta)\right) S_w(\theta)d\theta\leq P,\\
&\text{(strict causality constraints in frequency domain)}\\
&\int_{-\pi}^{\pi} a(\theta)\cos(n\theta) d\theta + \int_{-\pi}^{\pi} b(\theta) \sin(n\theta)d\theta = 0\\
& \quad n = 0, 1,2,\cdots,\infty \\
\end{split}
\label{formula_stationaryGuassian_upperbound_equi}
\end{equation}
where the maximum is taken over a functional set $\Gamma$ defined as
\begin{equation}
\begin{split}
\Gamma =& \lbrace a(\theta), b(\theta): [-\pi, \pi] \rightarrow \mathbb{R} \quad | \quad a(\theta), b(\theta)\in \mathcal{L}_2\rbrace.\\
\end{split}
\label{def_set_Gamma}
\end{equation}
\label{lemma:symmetric_filter}
\end{lemma}

With a bit abuse of notation, $S_w(\theta)$ refers to $\mathbb{S}_w(e^{j\theta})$ for simplicity.
The basic idea of this characterization is that the strict causality can be imposed on the non-positive index coefficients of the inverse Fourier transform of $\mathbb{Q}(e^{j\theta})$ by setting them to zeros. See Appendix for the detailed proof.

\subsection{Upper bounds}
We next obtain upper bounds to $C_{fb}$ by taking into account only a finite number of causality constraints. The $h$-upper-bound, denoted by $C_{fb}(h)$, is defined as
follows:
\begin{lemma}
Under Assumption \ref{LTIFD.ass} and \ref{nonwhite.ass}, define $C_{fb}(h)$ with $h \in \mathbb{Z}_+$ by
\begin{equation}\label{cfbh.eq}
\begin{split}
C_{fb}(h)=&\sup_{\Gamma}\frac{1}{4\pi}\int_{-\pi}^{\pi}\log ((1+a(\theta))^2+ b(\theta)^2 )d\theta\\
s.t. \quad &\frac{1}{2\pi}\int_{-\pi}^{\pi}\left(a^2(\theta)+b^2(\theta)\right) S_w(\theta)d\theta\leq P,\\
&\int_{-\pi}^{\pi} a(\theta)\cos(n\theta) d\theta + \int_{-\pi}^{\pi} b(\theta) \sin(n\theta)d\theta = 0\\
& \quad n = 0, 1,2,\cdots, h.\\
\end{split}
\end{equation}
Then, $C_{fb}(h) \geq C_{fb}(h+1), C_{fb}(h) \geq C_{fb}$ for any $h\geq 0$, and
$$C_{fb} = \lim_{h\rightarrow \infty} C_{fb}(h).$$
\label{lemma:C_h}
\end{lemma}
\begin{IEEEproof} $C_{fb}(h)$  is a monotonically non-increasing sequence bounded below, therefore it has a limit. The limit value must be $C_{fb}$, otherwise we can construct a strictly causal filter that achieves a rate strictly greater than the feedback capacity, which is impossible.
\end{IEEEproof}
The achievability of ``supremum'' will be proved in the next theorem.

Notice that $C_{fb}(h)$ is still an infinite dimensional problem. To solve $C_{fb}(h)$, we next provide a theorem that characterizes the Lagrangian dual of $C_{fb}(h)$ and show that there is no duality gap between the infinite dimensional primal and the finite dimensional dual problems. This result provides a convex optimization approach to compute $C_{fb}(h)$, a sequence of asymptotically capacity-achieving upper bounds.
\begin{theorem}(\textit{Main result})\label{strongdual.thm}
Under Assumption \ref{LTIFD.ass} and \ref{nonwhite.ass}, let
$$
\begin{array}{l}
A(\theta) =[\cos(\theta),\cos(2\theta), \cdots,\cos(h\theta)]',\\
B(\theta) =[\sin(\theta),\sin(2\theta), \cdots,\sin(h\theta)]'.
\end{array}
$$
For $\lambda > 0$, $\eta\in \mathbb{R}^h$, and $\eta_0\in \mathbb{R}$, define
$$
r^2(\theta) =(2\lambda S_{w}(\theta)+\eta'A(\theta)+\eta_0)^2+(\eta'B(\theta))^2.
$$
Then, the following statements are true.\\
a) The Lagrangian dual of (\ref{cfbh.eq}) is given by
\begin{equation}\label{dual.eq}
(D): \mu_h = - \max_{\lambda > 0,\eta\in \mathbb{R}^{h},\eta_0\in \mathbb{R}}g(\lambda,\eta,\eta_0)
\end{equation}
where
\begin{equation}\label{thm.dual.eq}
\begin{split}
&g(\lambda,\eta,\eta_0)= \displaystyle\frac{1}{2\pi}\int_{-\pi}^\pi \left[\frac{1}{2}\log(2\lambda S_{w}(\theta)-\nu(\theta))-\frac{r^2(\theta)}{2\nu(\theta)}+\lambda S_{w}(\theta)\right]d\theta -\lambda P+\eta_0+\frac{1}{2}.\\
\end{split}
\end{equation}
with
\begin{equation}\label{optimal.nu}
\nu(\theta)=\frac{-r^2(\theta)+\sqrt{r^4(\theta)+8\lambda S_{w}(\theta)r^2(\theta)}}{2}.
\end{equation}
\\
b) The dual problem (D) in (\ref{dual.eq}) is equivalent to the following convex optimization problem
\begin{equation}\label{dual2.eq}
\mu_h = - \max_{\begin{array}{l}\lambda > 0,\eta\in \mathbb{R}^{h},\eta_0\in \mathbb{R}\\ \nu(\theta)\geq 0 \in C^\infty_{[-\pi,\pi]}\end{array}}
\tilde{g}(\lambda,\eta,\eta_0,\nu(\theta))
\end{equation}
where
\begin{equation}\label{thm.dual2.eq}
\begin{split}
&\tilde{g}(\lambda,\eta,\eta_0,\nu(\theta)) =\displaystyle\frac{1}{2\pi}\int_{-\pi}^\pi \left[\frac{1}{2}\log(2\lambda S_{w}(\theta)-\nu(\theta))-\frac{r^2(\theta)}{2\nu(\theta)}+\lambda S_{w}(\theta)\right]d\theta -\lambda P+\eta_0+\frac{1}{2},\\
\end{split}
\end{equation}
and the optimal $\nu(\theta)$ is characterized by (\ref{optimal.nu}).\\
c) Furthermore, $C_{fb}(h)=\mu_h$, and an optimal filter $\mathbb{Q}_h(e^{j\theta}) = a(\theta)+jb(\theta)$ for $C_{fb}(h)$ exists and is characterized by
\begin{equation}\label{thm_xy_dual}
\begin{split}
a(\theta)=&\frac{2\lambda S_{w}(\theta)+\eta'A(\theta)+\eta_0}{\nu(\theta)}-1 \quad a.e.\\
b(\theta)=&\frac{\eta'B(\theta)}{\nu(\theta)}  \quad a.e.\\
\end{split}
\end{equation}
\label{lemma.strong.dual}
where $(\lambda, \eta', \eta_0, \nu(\theta))$ are obtained by solving (\ref{dual.eq}) (or (\ref{dual2.eq})).
\end{theorem}
%
\subsection{Computing $C_{fb}(h)$}
\indent Although the dual problem of $C_{fb}(h)$ can be cast into a convex
optimization (\ref{dual.eq}) with finite number of variables. The problem is not easily computable since the cost is an  integral, not explicitly computable in terms of the variables.
A natural practical approach would be to approximate the integral with a finite sum by discretizing $\theta$. We apply such discretization to (\ref{dual2.eq}) (with spacing $\frac{\pi}{m}$) and introduce the following finite dimensional convex problem. Given $m$,  consider
\begin{equation}{\label{opt_upperbound_approximate}}
\mu_h(m)=-\max_{\lambda > 0,\eta\in \mathbb{R}^{h},\eta_0\in \mathbb{R}, \nu_i\geq 0}\tilde{g}_m(\lambda,\eta,\eta_0,\nu_i)
\end{equation}
where
\begin{equation*}
\begin{split}
&\tilde{g}_m(\lambda,\eta,\eta_0,\nu_i)=  \frac{1}{2m}\sum_{i=1}^{2m}\left(\frac{1}{2}\log(2\lambda S_{w}(\theta_i)-\nu_i)+\lambda S_{w}(\theta_i)-\frac{r^2(\theta_i)}{2\nu_i} -\lambda P+\eta_0+\frac{1}{2}\right),\\
\end{split}
\end{equation*}
and $\theta_i = -\pi+ \frac{\pi}{m}(i-1)$. \\
{

Under Assumption \ref{ass0.ass} and \ref{LTIFD.ass} on $S_{w}(\theta)$, we know that for any given feasible ($\lambda,\eta,\eta_0,\nu(\theta)$) in (\ref{thm.dual2.eq}) the integrand function, defined in a compact set $[-\pi,\pi]$, is bounded (from Assumption \ref{ass0.ass}) and continuous (from Assumption \ref{LTIFD.ass}) almost everywhere\footnote{Discontinuity may exist when $\nu(\theta)=0$, but has zero measure. a fact that has been proved in the arguments before (\ref{x.eq})}.  This implies the Lebesgue's criterion for integrablility holds (or equivalently, Riemann integral holds). Thus, $\mu_h(m)$ is an approximation of $\mu_h$ and $\lim_{m\rightarrow \infty} \mu_h(m) = \mu_h$. }

Notice that the optimization (\ref{opt_upperbound_approximate}) is in a simple convex form. In particular, the $\log$ of an affine function is concave. $\frac{r^2(\theta_i)}{\nu_i}$ is a quadratic (composed with an affine function of the variables)  over linear function, therefore convex.  Thus,  (\ref{opt_upperbound_approximate}) can be efficiently  solved  with standard convex optimization tools, e.g. CVX, {a package for specifying and solving convex programs\cite{cvx01, cvx02}}.

Based on the solution to (\ref{opt_upperbound_approximate}), we can actually obtain a guaranteed upper bound on $C_{fb}(h)$ for each $m$ using the upper bound property of dual feasible solutions.
Let $\lambda ^{(m)}, \eta^{(m)}, \eta_0^{(m)}, \nu_i^{(m)}$ be the optimal solution to  (\ref{opt_upperbound_approximate}). Let
\begin{equation}{\label{C_h_m}}
\overline{C_{fb}(m,h)} = - g(\lambda ^{(m)}, \eta^{(m)}, \eta_0^{(m)})
\end{equation}
where $g(\cdot)$ is defined in (\ref{thm.dual.eq}). Clearly,  $\overline{C_{fb}(m,h)}$ is computable to  arbitrary accuracy.

\begin{corollary} \label{coro:upper_bound}
Given $h\geq 0$, $\overline{C_{fb}(m,h)} \geq C_{fb}(h) \geq C_{fb}$ for $\forall m > 0 $ and
$$C_{fb} = \lim_{h\rightarrow\infty}{C_{fb}(h)} = \lim_{h\rightarrow\infty}\lim_{m\rightarrow\infty}\overline{C_{fb}(m,h)}.$$
\end{corollary}

\section{Lower Bounds on Capacity}\label{sec::lowerbound}
In the previous section we have introduced a finite dimensional convex optimization (\ref{opt_upperbound_approximate}). From its optimal cost we were able to obtain a sequence of convergent upper bounds on $C_{fb}$. In this section, we show that from the solution to (\ref{opt_upperbound_approximate})  we can obtain lower bounds on $C_{fb}$ by explicitly constructing feedback codes. We show that these codes provide  lower bounds arbitrarily close to the capacity, providing a capacity-achieving feedback code.

The results of this section are summarized as follows.\\

{\bf Constructing Feedback Codes}
\begin{enumerate}
\item \textit{Filter Construction}:\\
 Given $h \in \mathbb{Z}_+$ and $m \in \mathbb{Z}_+ $ with $2m > h$. Solve (\ref{opt_upperbound_approximate}) to obtain solution $(\lambda_{h,m}, \eta_{h,m},\eta_{0,h,m}, \nu_{i,h,m})$.

For $i=1,\ldots,2m$ and $\theta_i = -\pi+ \frac{\pi}{m}(i-1)$, if $\nu_{i,h,m}>0$  compute $a_i$ and $b_i$,  from
\begin{equation}\label{thm_xy_dual_approx}
\begin{split}
a_i&=\frac{2\lambda_{h,m} \mathbb{S}_{w}(\theta_i)+\eta_{h,m}'A(\theta_i)+\eta_{0,h,m}}{\nu_{i,h,m}}-1 \\
b_i&=\frac{\eta_{h,m}'B(\theta_i)}{\nu_{i,h,m}}.\\
\end{split}
\end{equation}
If $\nu_{i,h,m}=0$\footnote{This case is not expected to occur for large enough $m$.} for some $i$'s,
$a_i,b_i$ can be obtained by solving (\ref{cfbhfd5.eq}), the dual problem of (\ref{opt_upperbound_approximate}).\footnote{We can also  compute $a_j$, $b_j$  by completion as they together with (\ref{thm_xy_dual_approx}) need to satisfy the causality constraints.
$$
\frac{1}{2m}\sum_{i=1}^{2m}a_i\cos(n\theta_i) +\frac{1}{2m} \sum_{i=1}^{2m} b_i \sin(n\theta_i) = 0, \quad n = 0, 1,\cdots, h.
$$}

Then construct a strictly causal filter $\mathbb{Q}_{N}^m(z)=\displaystyle\sum_{n=1}^N q_n^m z^{-n}$ {where $N=2m-h-1$ and}
\begin{equation}
\begin{array}{l}
q_n^m =\displaystyle\frac{1}{2m}\sum_{i=1}^{2m}a_i\cos(n\theta_i)-b_i\sin(n\theta_i).\\
\end{array}
\label{alg:L2_filter}
\end{equation}

\item \textit{Power Scale}:\\
 Let
 $$p:=\frac{1}{2\pi}\int_{-\pi}^{\pi}|\mathbb{Q}_{N}^m(e^{j\theta})|^2\mathbb{S}_w(e^{j\theta})d\theta.$$
Scale $\mathbb{Q}_{N}^m$ by $\alpha_{m,N}=\sqrt{P/p}$, i.e., $\mathbb{\bar{Q}}_{N}^m=\alpha_{m,N}\mathbb{Q}_{N}^m$, to satisfy the power budget $P$.
\item \textit{Coding Scheme Construction}: Construct a feedback coding scheme as described in Section \ref{sec.interpretation}.A by transforming $\mathbb{K} = \mathbb{\bar{Q}}_{N}^m(1+\mathbb{\bar{Q}}_{N}^m)^{-1}$ into the state-space representation.\footnote{Technically, Hankel Singular Value decomposition (H-SVD) procedure \cite{Hsvd} can be applied to arbitrarily well approximate an exact state-space realization of an FIR filter. As widely used to reduce the order of a system realization by such an approximation, one can use H-SVD to construct a low-order state-space representation of $\mathbb{\bar{Q}}_{N}^m$. }
\end{enumerate}
{
\begin{lemma}
Under Assumptions \ref{LTIFD.ass} and \ref{nonwhite.ass}, and given $m,h \in \mathbb{Z}_+$ with $2m > h$ and $N=2m-h-1$, the above coding scheme achieves a rate (in the sense of Shannon)
$$
R_N(m)= \frac{1}{2\pi}\int_{-\pi}^{\pi}\log |1+\mathbb{\bar{Q}}_{N}^m(e^{j\theta})|d\theta.
$$
\end{lemma}
This result directly follows from Theorem \ref{thm_capacity_achieving_code}. }The next theorem shows that the above constructed feedback code achieves capacity for sufficiently large $N$ and $m$.

\begin{theorem}\label{thm:lower_bound_convergence}
Under Assumptions \ref{LTIFD.ass} and \ref{nonwhite.ass}, and given $h, m \in \mathbb{Z}_+ $ with $2m > h$,
$$ \lim_{N \rightarrow \infty} \lim_{m \rightarrow \infty}R_N(m)= C_{fb}.$$
\end{theorem}
The proof is rather technical and thus presented in \ref{appendix: proof_LB}.
{The result of the theorem says that for $m$ and $N$ large enough there are codes whose rates are arbitrarily close to the feedback capacity. Equivalently, this implies that we can find $m$ and $h=2m-1-N$ so that the resulting code achieves a rate arbitrarily close to the feedback capacity.}


\section{Numerical Examples}
{ In this section, we provide examples to verify our results. We first summarize the procedure of computing the feedback capacity and constructing the capacity-achieving codes as follows. Given a finite dimensional Gaussian channel, we can compute a capacity upper bound $\overline{C_{fb}(m,h)}$ by solving the finite dimensional convex optimization (\ref{opt_upperbound_approximate}). Then we construct a feedback coding scheme as described in Section \ref{sec::lowerbound}. One will see that for the examples the capacity can be evaluated with arbitrary accuracy and the capacity-achieving code can be constructed explicitly.}

\subsection{First-order moving average Gaussian process}
Consider a first-order moving average (i.e. MA(1)) Gaussian process $W_i = U_i + 0.4U_{i-1}$ for $i\in \mathbb{Z}$, where $U_i$ is a white Gaussian process with zero mean and unit variance. The power spectral density is $\mathbb{S}_w(e^{j\theta}) = |1+ 0.4 e^{-j\theta}|^2$. Given power constraint $P = 10$, our proposed upper
and lower bounds converge to $1.8819$ \textit{bits/channel use}. This is consistent with that computed from the closed form solution (44) in \cite{Kim10}. { In particular, for a stationary noise process given by
$$W_i+\beta W_{i-1} = U_i + \alpha U_{i-1},$$ the feedback capacity is
\begin{equation*}
C_{fb} = - \log x_0,
\end{equation*}
where $x_0$ is the unique positive root of the fourth-order polynomial in $x$,
$$ Px^2 = \frac{(1-x^2)(1+\sigma\alpha x)^2}{(1+\sigma\beta x)},$$
and
$$ \sigma = sign(\beta - \alpha).$$

To obtain a quantitative flavor of the capacity gain of using feedback, we compute the non-feedback capacity as a comparison, according to the formula (2), (3) in \cite{Kim10}. It turns out that the non-feedback capacity for this channel is $1.7402$ \textit{bits/channel use}. That is, the capacity gain is $0.1417$ \textit{bits/channel use} by using feedback.}

\subsection{Second-order moving average Gaussian process}
Consider the following second order moving average Gaussian process with
$$
\mathbb{H}(z)=1+0.1z^{-1}+0.5z^{-2}
$$
with associated $\mathbb{S}_w(e^{j\theta})=|\mathbb{H}(e^{j\theta})|^2$. While neither the value of capacity or the optimal code are known for this generalized Gaussian noise, both of them can be efficiently obtained from our approach.
With power constraint $P = 10$, the capacity is evaluated to be $1.9194$ \textit{bits/channel use} (rounded to $4$ decimals). { Fig. \ref{gap.fig} shows the exponentially fast convergence of the upper and lower bounds in this example. Furthermore, we note that the non-feedback capacity is $1.7466$ \textit{bits/channel use}, from which we see the noticeable capacity gain by using feedback.
\begin{figure}
\begin{center}
\includegraphics[scale=.6]{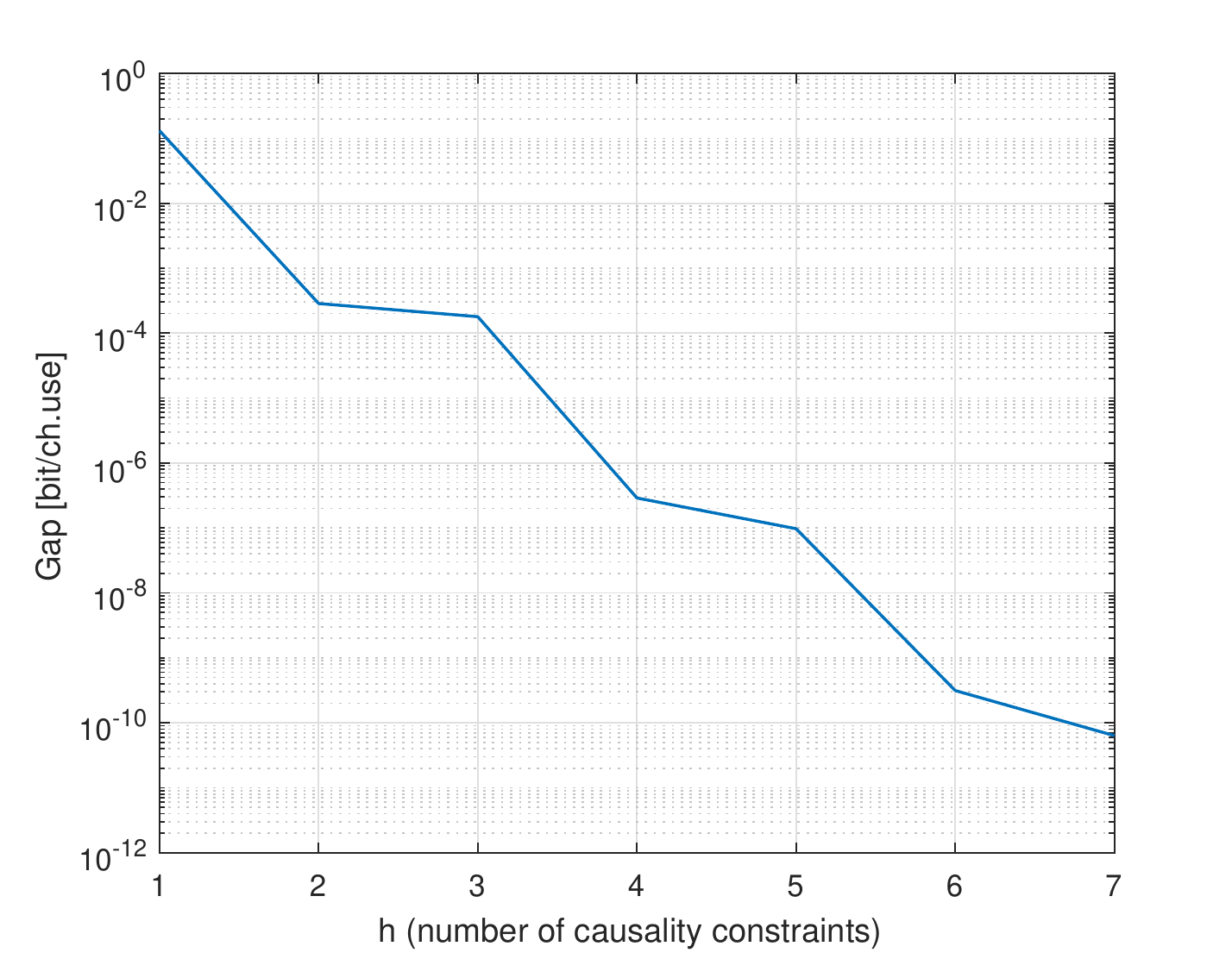}
\caption{Gap between upper and lower bounds exponentially vanishes as the number of causality constraints increases.}
\label{gap.fig}
\end{center}
\end{figure}

The optimal feedback coding scheme $\mathbb{K}$  after order reduction via H-SVD on the finite impulse response is given by
$$
{\mathbb K}= \frac{0.22026 (z+13.84) z^2}{(z^2 + 0.01755z + 0.03498) (z^2 + 0.4115z + 3.783)}.
$$
Then, we can obtain the corresponding state-space representation in the real block diagonal form and construct the explicit coding scheme as shown in Fig. \ref{fig:codingStructure}.
\begin{figure}
\begin{center}
\includegraphics[scale=.65]{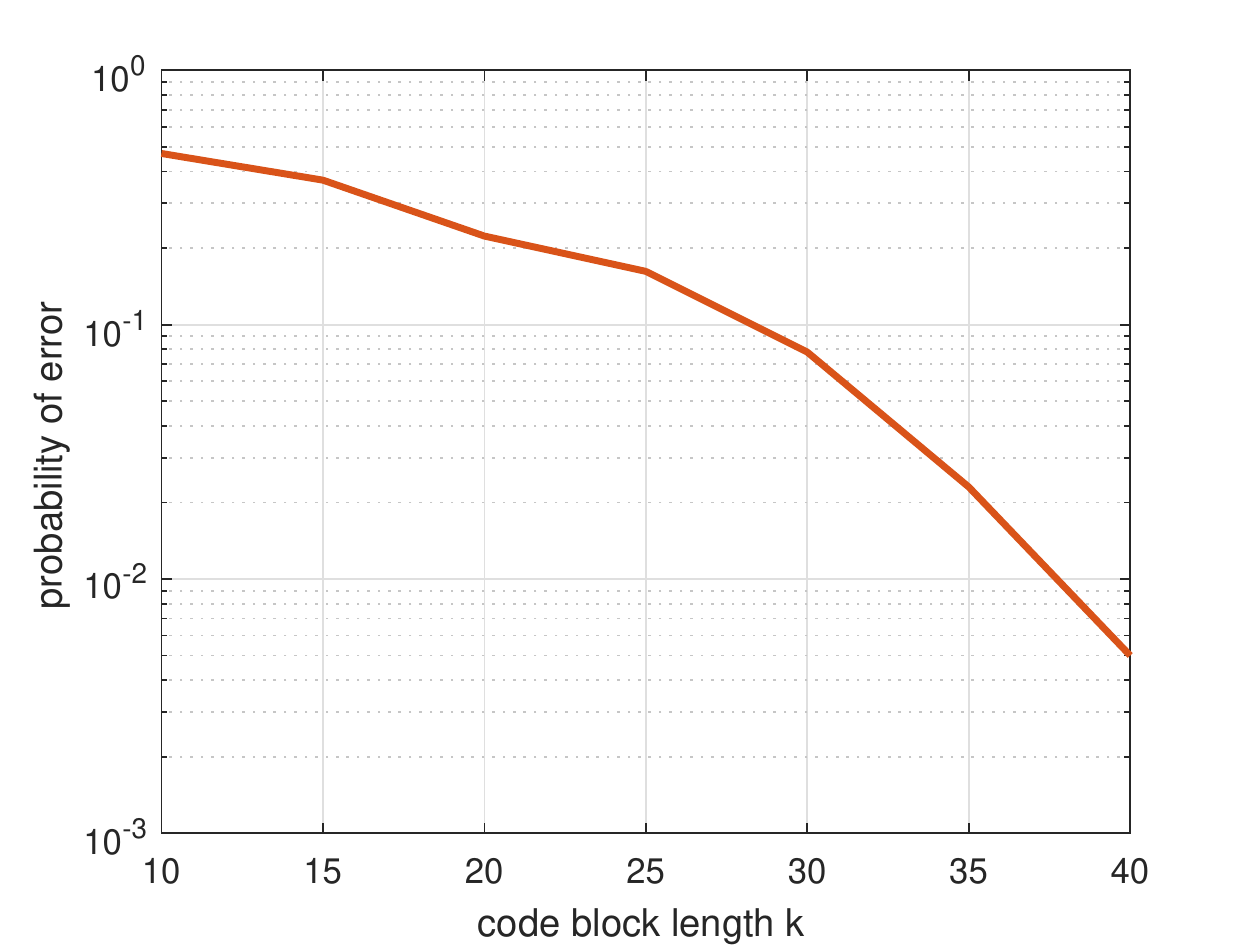}
\caption{Decoding error probability at rate R = $0.95 C_{fb}$.}
\label{code_perf.fig}
\end{center}
\end{figure}
Note that $\mathbb{K}$ has two complex conjugate unstable poles at
$$
p_{1,2}=-0.2057 \pm i 1.9340,
$$
which would not be easy to find using the approach of \cite{Elia2004}. According to Theorem \ref{thm_capacity_achieving_code}, we know the reliable transmission rate is only determined by the unstable poles of the constructed controller $\mathbb{K}$. Therefore, the fact that the above capacity-achieving coding scheme $\mathbb{K}$ is unstable is expected.  Also it can be verified from Theorem \ref{thm_capacity_achieving_code} that the achievable rate of $\mathbb{K}$ is $\log(|p_1||p_2|)=1.9194$ \textit{bits/channel use}.

Moveover, the corresponding optimal closed loop Sensitivity function is
$$
1+{\mathbb Q}=\frac{(z^2 + 0.01755z + 0.03498) (z^2 + 0.4115z + 3.783)}{(z^2 + 0.1088z + 0.2644) (z^2 + 0.1z + 0.5)}
$$
As expected, the Sensitivity has the corresponding non-minimum phase zeros at the location of the unstable poles of ${\mathbb K}$.
Note also that the optimal closed loop system includes dynamics that (partially) cancel the noise dynamics. The term $( z^2 + 0.1z + 0.5)$ is the numerator of $\mathbb{H}(z)$.
This feature is to be consistent with the optimal filter structure identified in \cite{Kim10}.

Fig. \ref{code_perf.fig} \footnote{The simulation procedure, including how to define the message index and determine the detection rule/threshold, can be found in the proof of Theorem 4.3 in \cite{Elia2004}.} shows that the error probability of the constructed code decays to zero in a double exponential manner. This implies the coding latency can be very small to achieve certain performance.

\subsection{Third-order ARMA Gaussian process}
Finally, Fig. \ref{Capacity_curve.fig} shows the feedback capacity curve as a function of power budget $P$ for an ARMA(3) noise with
\begin{equation}\label{arma3.eq}
\mathbb{H}(z)=\frac{z^3-0.3z^{2}+0.5z+0.2}{z^3+0.1z^2+0.6z+0.5}.
\end{equation}
It is noteworthy that the curve is achievable (within any epsilon) by the codes we have proposed and such curve was unknown until now.}
\begin{figure}
\begin{center}
\includegraphics[scale=.4]{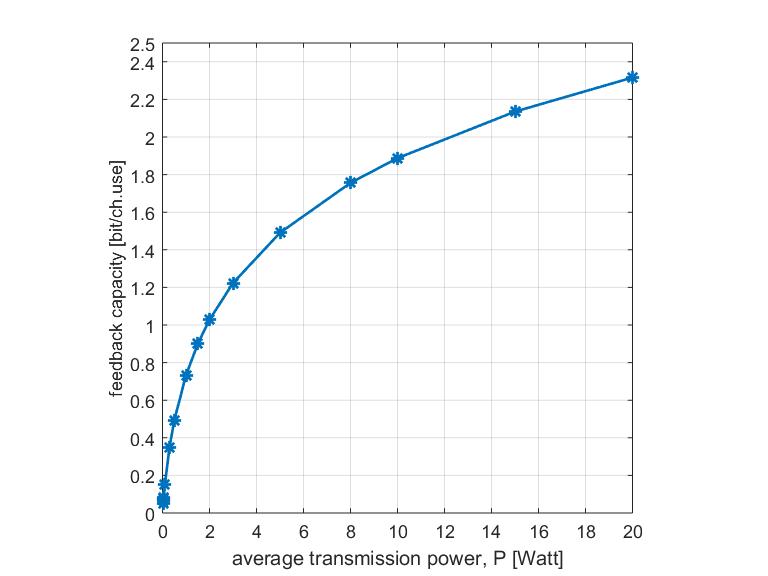}
\caption{Feedback capacity curve for the noise given by (\ref{arma3.eq}). }
\label{Capacity_curve.fig}
\end{center}
\end{figure}


%

\section{Conclusion}
This paper studied the problem of computing the feedback capacity of stationary finite dimensional Gaussian channels and found the asymptotically capacity-achieving codes. Firstly, the interpretation of feedback communication as feedback control over Gaussian channels was extended by leveraging \textit{Youla} parameterization. This new interpretation provides an approach to construct feasible feedback coding schemes with double exponentially decaying error probability. We next derived an asymptotic capacity-achieving upper bounds, which can be numerically computed by solving finite dimensional convex optimizations. From the resulting filters that achieve upper bounds, feasible feedback coding schemes were constructed, leading to a sequence of lower bounds. We proved that the sequence of lower bounds is asymptotically capacity-achieving.

{In summary, the paper provides a computational approach to compute the feedback capacity of Gaussian channels and construct the capacity-achieving feedback codes, both of which have been open problems in the literature. Furthermore, we verified our results by numerical examples. In particular, we computed the capacity and constructed the capacity-achieving feedback codes for a MA(2) channel, and also presented the capacity curve of an ARMA(3) channel. These results were not known/reported in the literature. Moreover, the resulting structure of the capacity-achieving feedback codes, obtained from our control-based numerical approach, could also provide insight for further investigation on the analytical/closed-form solutions on feedback capacity and capacity-achieving coding scheme.}

\section{Appendix}
\subsection{Proof of Lemma \ref{lem.direct.info.power}}
Denote $h(\cdot)$ as the differential entropy, then
$$
\begin{array}{l}
\displaystyle\stackrel{\rightarrow}{I}(U^T\rightarrow Y^T)\\
=\displaystyle\sum_{t=0}^TI(U^t;Y_t|Y^{t-1})\\
=\displaystyle\sum_{t=0}^Th(Y_t|Y^{t-1})-h(Y_t|Y^{t-1},U^t)\\
=\displaystyle\sum_{t=0}^Th(Y_t|Y^{t-1})-h(W_t+U_t|Y^{t-1},U^t)\\
=\displaystyle\sum_{t=0}^Th(Y_t|Y^{t-1})-h(W_t |W^{t-1},U^t)\\
=\displaystyle\sum_{t=0}^Th(Y_t|Y^{t-1})-h(W_t |W^{t-1})\\
=h(Y^T)-h(W^T)\\
=\displaystyle\frac{1}{2}\log \frac{Det [\Sigma_{Y^T}]}{Det[\Sigma_{W^T}]},
\end{array}
$$
where the last step follows from the fact that $Y^T, W^T$ are Gaussian vectors and $\Sigma_{Y^T}$ denotes the covariance matrix of sequence $Y^T$. Now, according to (\ref{Y.eq}) (similar to Theorem 4.6 in \cite{Elia2004}), it yields
$$
\begin{array}{l}
\displaystyle\lim_{T\to \infty}\frac{1}{T}\displaystyle\stackrel{\rightarrow}{I}(U^T\rightarrow Y^T)\\
=\displaystyle\lim_{T\to \infty}\frac{1}{2T}\log \frac{Det [\Sigma_{Y^T}]}{Det[\Sigma_{W^T}]}\\
=\displaystyle\frac{1}{2\pi}\int_{-\pi}^\pi\frac{1}{2}\log\frac{|\mathbb{Y}(e^{j\theta})|^2}{|\mathbb{W}(e^{j\theta})|^2}d\theta\\
=\displaystyle\frac{1}{2\pi}\int_{-\pi}^\pi\frac{1}{2}\log\frac{\mathbb{S}_Y(e^{j\theta})}{\mathbb{S}_w(e^{j\theta})}d\theta\\
=\displaystyle\frac{1}{4\pi}\int_{-\pi}^\pi\log\frac{|1+\mathbb{Q}(e^{j\theta})|^2\mathbb{S}_w(e^{j\theta})+|\mathbb{Q}_v(e^{j\theta})|^2\mathbb{S}_v(e^{j\theta}))}{\mathbb{S}_w(e^{j\theta})}d\theta\\
=\displaystyle\frac{1}{4\pi}\int_{-\pi}^\pi\log\left(|1+\mathbb{Q}(e^{j\theta})|^2+\frac{|\mathbb{Q}_v(e^{j\theta})|^2\mathbb{S}_v(e^{j\theta})}{\mathbb{S}_w(e^{j\theta})}\right)d\theta.\\
\end{array}
$$
The average power formula is derived as follows,
\begin{equation}
\begin{split}
&\lim_{T\rightarrow \infty} \frac{1}{T}\sum_{t=1}^{T}U_t^2\\
\stackrel{(a)}{=} &\frac{1}{2\pi}\int_{-\pi}^\pi |{\mathbb S}_u(e^{j\theta})|^2d\theta\\
\stackrel{(b)}{=} &\frac{1}{2\pi}\int_{-\pi}^\pi |\mathbb{Q}(e^{j\theta})|^2{\mathbb S}_w(e^{j\theta})+|\mathbb{Q}_v(e^{j\theta})|^2\mathbb{S}_v(e^{j\theta}) d\theta,\\
\end{split}
\end{equation}
where step (a) follows from Parseval's theorem and step (b) follows from (\ref{U.eq}) and the independence of $W$ and $V$.

%

\subsection{Proof of Lemma \ref{lemma_symmetric_filter}}

Without loss of generality, we assume a feasible $\mathbb{Q}(e^{j\theta})$ satisfying $\mathbb{Q}(e^{j\theta})=\mathbb{Q}^*(e^{-j\theta})$. Let $\mathbb{Q}(e^{j\theta}) = a(\theta) + j b(\theta) =\sum_{n = -\infty}^{\infty} c_n e^{-jn\theta}$   where $a(\theta) = \sum_{n = -\infty}^{\infty} c_n \cos(n\theta)$ and $b(\theta)=\sum_{n = -\infty}^{\infty} -c_n \sin(n\theta)$ represent the real and imaginary part of $\mathbb{Q}(e^{j\theta})$, respectively. The average power constraint implies $a(\theta),b(\theta) \in \mathcal{L}_2$. It is worth noting that the condition $\mathbb{Q}(e^{j\theta})=\mathbb{Q}^*(e^{-j\theta})$ indicates $\mathbb{Q}(e^{j\theta})$ is the Fourier transform of a real signal (impulse response of filter $\mathbb{Q}(e^{j\theta})$) \cite{MIT_course}. This implies
\begin{equation}
a(-\theta) = a(\theta), \quad b(-\theta) = -b(\theta).
\label{parameter_even_odd}
\end{equation}
\indent Next, a \textit{strictly causal} $\mathbb{Q}(e^{j\theta})$ implies the Fourier coefficients
\begin{equation}
c_n = \int_{-\pi}^{\pi}\mathbb{Q}(e^{j\theta})e^{jn\theta} \frac{d\theta}{2\pi}
\label{filter_coefficient}
\end{equation}
satisfy $c_n = 0$ for all $n\leq 0$. Specifically,
\begin{equation}
c_n = \int_{-\pi}^{\pi} (a(\theta) + j b(\theta))(\cos(n\theta) + j\sin(n\theta)) \frac{d\theta}{2\pi}= 0
\label{equ_fourier_coeff}
\end{equation}
for $n \leq 0$. After some elementary algebra and based on the fact that function $a(\theta)$ is even and function $b(\theta)$ is odd, we have
\begin{equation}
\begin{split}
c_n = 0 \Leftrightarrow \int_{-\pi}^{\pi} a(\theta)\cos(n\theta)d\theta - \int_{-\pi}^{\pi} b(\theta)\sin(n\theta)d\theta = 0.\\
\end{split}
\label{equ.coeff.equality01}
\end{equation}
Now we change the sign of index $n$ and have
\begin{equation}
c_{-n} = \int_{-\pi}^{\pi} a(\theta)\cos(n\theta)\theta + \int_{-\pi}^{\pi} b(\theta)\sin(n\theta)d\theta = 0\\
\label{equ_coeff_equality}
\end{equation}
for $n \geq 0$.
The proof is complete.

\subsection{Proof of Theorem \ref{lemma.strong.dual}}
{ To facilitate the technical reading, we first provide the high-level discussion on the proof as follows. To derive the characterization of the dual function, we leverage the optimality conditions when minimizing the Lagrangian function. These conditions are presented in the form of equations linking the primal and dual variables, based on which we successfully characterize the minimized Lagrangian function, or equivalently, the dual function. Next, to show the strong duality, we explicitly construct a primal solution to the primal problem based on the selected dual variables. Then we show that the dual objective value based on the selected dual variables and the primal objective value based on the constructed primal solution coincide.}

For the ease of derivation, we first change some notations in $C_{fb}(h)$. Let $x(\theta) = a(\theta) + 1$, $y(\theta) = b(\theta)$. Denote $A(\theta) = [\cos(\theta),\cos(2\theta), \cdots,\cos(h\theta)]' $, and
$B(\theta) = [\sin(\theta),\sin(2\theta), \cdots,\sin(h\theta)]' $. It is true that $A(\theta)$ and $B(\theta)$ have  full row rank and moreover the row of $A(\theta)$ and $B(\theta)$ are orthonormal. Consider
\begin{equation}
\begin{array}{ll}
(P): &\mu_h^P= \inf_{x,y,c}\displaystyle\frac{1}{4\pi}\int_{-\pi}^\pi -\log(x^2(\theta)+y^2(\theta)+c^2(\theta))d\theta\\
& s.t. \\
&\displaystyle\frac{1}{2\pi}\int_{-\pi}^\pi S_{w}(\theta)(x^2(\theta)+y^2(\theta)+c^2(\theta)-2x(\theta)+1)d\theta\leq P,\\
&\displaystyle\frac{1}{2\pi}\int_{-\pi}^\pi(A(\theta)x(\theta)+B(\theta)y(\theta))d\theta=0,\\
&\displaystyle\frac{1}{2\pi}\int_{-\pi}^\pi x(\theta) d\theta=1, \quad {x(\theta),y(\theta),c(\theta)\in \mathcal{L}_2}.\\
\label{proof:strong_dual_primal_problem}
\end{array}
\end{equation}
We see that $C_{fb}(h)=-\mu_h^P$ if $c(\theta)=0$, a.e. {Alternatively, $\mu_h^P$ can be directly derived from (\ref{capacity_long01}). That is, letting $c^2(\theta)=\frac{\mathbb{S}_v(e^{j\theta)}}{\mathbb{S}_w(e^{j\theta})}$, $\mathbb{Q}(e^{j\theta}) = a(\theta) + j b(\theta)= x(\theta)-1 + j y(\theta)$ in (\ref{capacity_long01}), and following the derivation of Lemma \ref{lemma_symmetric_filter} by respectively replacing $a(\theta)$, $b(\theta)$ with $x(\theta)-1$ and $y(\theta)$, we can directly obtain the causality constraints (the last two equalities) in $\mu_h^P$. Then plugging $\mathbb{Q}(e^{j\theta}) = x(\theta)-1 + j y(\theta)$ into the objective function and the power constraint in (\ref{capacity_long01}) leads to the characterization on $\mu_h^P$.} As will be seen, we always have $c(\theta)=0$, a.e., for non-flat Gaussian noise.\\

\subsubsection{Derivation of dual problem}
$ $\\

Notice that the problem of $\mu_h^P$ is neither convex nor quasi-convex.  We next construct the dual and analyze optimality conditions. Consider the Lagrangian

\begin{equation}\label{lap.eq}
\begin{array}{ll}
&L(x,y,c, \lambda,\eta,\eta_0)\\
=&\displaystyle\frac{1}{2\pi}\int_{-\pi}^\pi \left[-\frac{1}{2}\log(x^2(\theta)+y^2(\theta)+c^2(\theta))\right.\\
&+ \lambda S_{w}(\theta)(x^2(\theta)+y^2(\theta)+c^2(\theta)-2x(\theta)+1)\\
&\left.-\eta'(A(\theta)x(\theta)+B(\theta)y(\theta))-\eta_0x(\theta)\right]d\theta -\lambda P +\eta_0,\\
\end{array}
\end{equation}
where $\lambda\geq 0, \eta\in \mathbb{R}^{h}, \eta_0\in \mathbb{R}$. The Lagrangian dual function is
\begin{equation}\label{ldf.eq}
g(\lambda,\eta,\eta_0)=\inf_{x,y,c} L(x,y,c, \lambda,\eta,\eta_0).
\end{equation}
Note that for any $\lambda\geq 0, \eta\in \mathbb{R}^{h}, \eta_0$, $g(\lambda,\eta,\eta_0)$ provides a lower bound to $\mu_h^P$. The optimal dual objective value is given as follows,
\begin{equation}\label{optmuD.eq}
 \mu_h^D= \sup_{\lambda\geq 0,\eta,\eta_0}g(\lambda,\eta,\eta_0).
\end{equation}
In sequel, we use ``$L$'' to represent ``$L(x,y,c,\lambda,\eta,\eta_0)$'' unless specified otherwise. In what follows, we derive the dual function $g(\lambda,\eta,\eta_0)$ by leveraging the optimality conditions.

First of all, the optimality conditions for $g$ are obtained by taking functional derivative\footnote{Note that when the space of functions is defined in a Banach space, the functional derivative becomes known as the Fr\'echet derivative; when the space of functions is defined in a more general locally convex space, one uses the G\^ateaux derivative \cite{Frechet_derivative, book.opt.vector.space}.} of $ L(x,y, c,\lambda,\eta, \eta_0)$ with respect to function $x$, $y$ and $c$. In particular,
\begin{equation}\label{primal_optimal_condition}
\begin{array}{l}
(x): \frac{x(\theta)}{x^2(\theta)+y^2(\theta)+c^2(\theta)}=2\lambda S_{w}(\theta)x(\theta)-2\lambda S_{w}(\theta)-\eta'A(\theta)-\eta_0\\
(y): \frac{y(\theta)}{x^2(\theta)+y^2(\theta)+c^2(\theta)}=2\lambda S_{w}(\theta)y(\theta)-\eta'B(\theta)\\
(c): \frac{c(\theta)}{x^2(\theta)+y^2(\theta)+c^2(\theta)}=2\lambda S_{w}(\theta)c(\theta).
\end{array}
\end{equation}

Note that substituting these conditions back into (\ref{lap.eq}) (hint: taking $(x)\cdot x + (y) \cdot y + (c)\cdot c$),
\begin{equation}\label{lapmin.eq}
\begin{array}{ll}
L=&\displaystyle\frac{1}{2\pi}\int_{-\pi}^\pi \left[-\frac{1}{2}\log(x^2(\theta)+y^2(\theta)+c^2(\theta))+\lambda S_w(\theta)\right.\\
&\left.- \lambda S_{w}(\theta)(x^2(\theta)+y^2(\theta)+c^2(\theta))\right]d\theta -\lambda P +\eta_0+1,\\
\end{array}
\end{equation}
which is monotonically decreasing with $(x^2(\theta)+y^2(\theta)+c^2(\theta))$.

Let
$$
\nu(\theta)=-\frac{1}{x^2(\theta)+y^2(\theta)+c^2(\theta)}+2\lambda S_{w}(\theta).
$$
Then \begin{equation}\label{nu.eq}
(x^2(\theta)+y^2(\theta)+c^2(\theta))=\frac{1}{2\lambda S_{w}(\theta)-\nu(\theta)}, \quad 2\lambda S_{w}(\theta)>\nu(\theta).
\end{equation}

%

Furthermore, the optimality conditions (\ref{primal_optimal_condition}) imply that the following equalities need to be satisfied
\begin{equation}\label{optcond.eq}
\begin{array}{l}
 \nu(\theta)x(\theta)=2\lambda S_{w}(\theta)+\eta'A(\theta)+\eta_0,\\
 \nu(\theta)y(\theta)=\eta'B(\theta),\\
 \nu(\theta)c(\theta)=0.
\end{array}
\end{equation}
For further use, the above assignments yield that
\begin{equation}\label{reducex.eq}
\begin{array}{ll}
&-(2\lambda S_{w}(\theta)+\eta'A(\theta)+\eta_0)x(\theta)+\frac{\nu(\theta)x^2(\theta)}{2}=-\frac{2\lambda S_{w}(\theta)+\eta'A(\theta)+\eta_0}{2}x(\theta),\\
\end{array}
\end{equation}
and
\begin{equation}\label{reducey.eq}
-\eta'B(\theta)y(\theta)+\frac{\nu(\theta) y^2(\theta)}{2}=-\frac{\eta'B(\theta)}{2}y(\theta).
\end{equation}
Squaring and summing both sides of (\ref{optcond.eq}), we have
\begin{equation}\label{rsquare.eq}
\nu(\theta)^2(x^2(\theta)+y^2(\theta)+c^2(\theta))=r^2(\theta),
\end{equation}
where
\begin{equation}\label{rsquare.eq_01}
r^2(\theta) = (2\lambda S_{w}(\theta)+\eta'A(\theta)+\eta_0)^2+(\eta'B(\theta))^2, \; \forall \theta\in[-\pi,\pi].
\end{equation}
Substituting $(x(\theta)^2+y(\theta)^2+c^2(\theta))$ from (\ref{nu.eq}) and rearranging the equation, we have
\begin{equation}\label{nu_2ord.eq}
r^2(\theta)(2\lambda S_{w}(\theta)-\nu(\theta))=\nu^2(\theta),
\end{equation}
which is a second order equation in $\nu(\theta)$ (for each $\theta$) that depends on the dual variables $\lambda, \eta, \eta_0$. Solving for $\nu(\theta)$ we have
$$
\nu(\theta)_{1,2}=\frac{-r^2(\theta)\pm\sqrt{r^4(\theta)+8\lambda S_{w}(\theta)r^2(\theta)}}{2}.
$$
From (\ref{nu.eq}), we notice that a larger $\nu(\theta)$ leads to a greater $(x^2(\theta)+y^2(\theta)+c^2(\theta))$ that decreases $L(x,y,c,\lambda,\eta,\eta_0)$ in (\ref{lapmin.eq}).
Therefore, since $\lambda\geq 0$, $\nu(\theta)\geq 0$ can always be found and we can concentrate on the positive solution
\begin{equation}\label{nucf.eq}
\nu(\theta)=\frac{-r^2(\theta)+\sqrt{r^4(\theta)+8\lambda S_{w}(\theta)r^2(\theta)}}{2}.
\end{equation}
The above derivations show that the desired $\nu(\theta)$ with $0\leq \nu(\theta)<2\lambda S_{w}(\theta)$ can always be found.
Here it is true that $\lambda>0$. This is because based on (\ref{lapmin.eq}) the case of $\lambda=0$ can be trivially ruled out as $L(x,y,c,\lambda,\eta,\eta_0)\rightarrow -\infty$ by increasing $x^2(\theta)+y^2(\theta)+c^2(\theta)$.

We next remark that,  $\nu(\theta)=0$ for only finite number of $\theta$. The reason is given as follows. Given $\lambda>0$, it is true that $\nu(\theta)=0$ for $\theta=\bar{\theta}$ if and only if $r(\bar{\theta})=0$.
If $r^2(\theta)=0$ for an infinite number of $\theta_i$, according to (\ref{rsquare.eq}), since $B(\theta)$ is full finite column rank, we have $\eta=0$, then $2\lambda S_w(\theta_i)+\eta_0=0$. However, since $S_w(\theta)= |\mathbb{H}(e^{j\theta})|^2$ is the spectral density of the output of a finite dimensional LTI stable system driven by white noise, it is a rational function of $\cos(\theta)$. Therefore
there must exist a finite number of roots for the equation $2\lambda S_w(\theta_i)+\eta_0=0$, unless $S_w(\theta)$ is constant, which is excluded by Assumption \ref{nonwhite.ass}.

Based on this discussion, $x(\theta)$, $y(\theta)$  and $c(\theta)$ satisfying the optimality conditions  can be obtained from (\ref{optcond.eq}) for almost all $\theta$, namely,
\begin{equation}\label{x.eq}
x(\theta)=\frac{2\lambda S_{w}(\theta)+\eta'A(\theta)+\eta_0}{\nu(\theta)}, \quad a.e.
\end{equation}
\begin{equation}\label{y.eq}
y(\theta)=\frac{\eta'B(\theta)}{\nu(\theta)}, \quad a.e.
\end{equation}
\begin{equation}\label{c.eq}
c(\theta)=0, \quad a.e.
\end{equation}
At those finite number of $\theta$'s, if any, where $\nu(\theta)=0$, $x$ and $y$ may not be well-defined and can be discontinuous, however, these points have zero measure. Furthermore, one can check that, under the above assignments of $x(\theta)$, $y(\theta)$ and $c(\theta)$  with $\nu(\theta)$ in (\ref{nucf.eq}), the optimality conditions (\ref{primal_optimal_condition}) for the dual function $g(\lambda, \eta,\eta_0)$ are satisfied.

Note that the above analysis shows that for non-flat channels, the optimal solution must have the feedforward component $c(\theta)=0$ (i.e. ${\mathbb S}_v(\theta)=0$ a.e.) and therefore the only contribution to the communication rate is obtained by feedback.
From (\ref{rsquare.eq}) and (\ref{nu_2ord.eq}), we have
\begin{equation}\label{rsquare.eq_ext}
\lambda S_{w}(x^2(\theta)+y^2(\theta)+c^2(\theta))=\frac{1}{2}+\frac{r^2(\theta)}{2\nu(\theta)}.
\end{equation}
Plugging (\ref{rsquare.eq}), (\ref{nu_2ord.eq}) and (\ref{rsquare.eq_ext}) into the Lagrangian,
the Lagrangian Dual function $g(\lambda,\eta,\eta_0)$ is characterized by
{\small{\begin{equation}\label{Lagdual_b.eq}
\begin{split}
&g(\lambda,\eta,\eta_0)=\frac{1}{2\pi}\hspace{-1.5mm}{\mathop{\mathlarger{\mathlarger{\mathlarger{\int}}}}}_{\hspace{-3mm}-\pi}^\pi
\left[\frac{1}{2}\log(2\lambda S_{w}(\theta)\hspace{-.5mm}-\hspace{-.5mm}\nu(\theta))\hspace{-.5mm}-\hspace{-.5mm}\frac{r^2(\theta)}{2\nu(\theta)}\hspace{-.5mm}+\hspace{-.5mm}\lambda S_{w}(\theta) \right] d\theta \hspace{-.5mm}-\hspace{-.5mm}\lambda P\hspace{-.5mm}+\hspace{-.5mm}\eta_0\hspace{-.5mm}+\hspace{-.5mm}\frac{1}{2},\\
\end{split}
\end{equation}}}
and
\begin{equation}\label{Lagdual_c.eq}
\mu_h^D\hspace{-.5mm}=\hspace{-2.5mm}\displaystyle\max_{\hspace{-2mm}\lambda > 0, \eta, \eta_0}\displaystyle\hspace{-1mm}g(\lambda,\eta,\eta_0),
\end{equation}
where $r^2(\theta)$ is given by (\ref{rsquare.eq_01}) and $\nu(\theta)$ given by (\ref{nucf.eq}) is a continuous bounded function on a closed set $\theta\in [-\pi,\pi]$, i.e., $\nu(\theta) \in C^{\infty}_{[-\pi,\pi]}$. Notice that ``$\sup$'' is replaced by ``$\max$'' due to the existence of solutions in $g(\lambda,\eta,\eta_0)$, which is proved in the lemma below.\\

\begin{lemma}\label{lemma.exist}
Under Assumption \ref{LTIFD.ass} and Assumption \ref{nonwhite.ass} (i.e. $c(\theta) = 0,$ a.e.), an optimal bounded solution $(\lambda,\eta,\eta_0)$ to $\mu_h^D$ exists.\\
\end{lemma}

The proof is given in Appendix \ref{proof:existence}. As a result, the characterization of dual problem (\ref{dual.eq}) completes. Notice the sign change $\mu_h=-\mu_h^D$ due to the assumption of $C_{fb}(h)=-\mu_h^P$ at the beginning of this proof.

The above optimization problem has finite number of variables, however, the cost is an integral function of  the dual variables.  To simplify the computation of the dual optimization, we introduce a relaxation of (\ref{Lagdual_b.eq}) by treating function $\nu(\theta)$ as a free variable, namely,
{\small{\begin{equation}\label{Lagdual.eq}
\tilde{\mu}_h^D\hspace{-.5mm}=\hspace{-4.5mm}\displaystyle\max_{\hspace{-2mm}{\scriptsize\begin{array}{l}\lambda > 0, \eta, \eta_0\\ \nu(\theta)\geq 0\\
\nu(\theta) \in {{C^\infty_{[-\pi,\pi]}}}\end{array}}}
\displaystyle\hspace{-3mm}\frac{1}{2\pi}\hspace{-2mm}{\mathop{\mathlarger{\mathlarger{\mathlarger{\int}}}}}_{\hspace{-3mm}-\pi}^\pi\hspace{-3mm}
\frac{\log(2\lambda S_{w}(\theta)\hspace{-.5mm}-\hspace{-.5mm}\nu(\theta))}{2}\hspace{-.5mm}-\hspace{-.5mm}\frac{r^2(\theta)}{2\nu(\theta)}\hspace{-.5mm}+\hspace{-.5mm}\lambda S_{w}(\theta) d\theta\hspace{-.5mm}-\hspace{-.5mm}\lambda P\hspace{-.5mm}+\hspace{-.5mm}\eta_0\hspace{-.5mm}+\hspace{-.5mm}\frac{1}{2}.
\end{equation}}}

We next show that this relaxation does not lose optimality, i.e., $\tilde{\mu}^D_h = \mu^D_h$. Firstly, it is true that $\tilde{\mu}^D_h\geq \mu^D_h$ as the problem of $\tilde{\mu}^D_h$ is less constrained. { To optimize $\tilde{\mu}^D_h$, we take gradients over $\lambda, \eta, \eta_0$ and functional derivative w.r.t. function $\nu(\theta)$ and equate them to zero. This leads to the equations in (\ref{max_condition}).
Using Leibniz's rule we can obtain the gradient w.r.t. to $\lambda, \eta, \eta_0$. For the functional derivative of the objective function of $\tilde{\mu}^D_h$ w.r.t. to $\nu(\theta)$, note that the integral is minimized w.r.t. $\nu$ if the integrand (denoted by $\tilde{g}$) is minimized w.r.t. $\nu(\theta)$ for all $\theta$, leading to (\ref{derivative_nu}) (\ref{max_condition}). Note that the last condition in (\ref{max_condition}) corresponds to (\ref{nu_2ord.eq}) and thus (\ref{nucf.eq}). This guarantees that $\nu(\theta)$ is a continuous function. } In addition, this indicates that given feasible $\lambda > 0$, $\eta$ and $\eta_0$, the cost functions of the two optimization problems coincide (under optimized $\nu(\theta)$). Therefore, $\tilde{\mu}^D_h = \mu^D_h$.\\

%
%

{
\begin{figure*}
\begin{equation}
\begin{array}{ll}\label{derivative_nu}
&\displaystyle \frac{\partial \tilde{g}}{\partial \nu(\theta)} = \displaystyle \frac{1}{2\pi} \left(\frac{-1}{2(2\lambda S_{w}(\theta) -\nu(\theta))}+ \frac{r^2(\theta)}{2\nu^2(\theta)}\right) = 0, \\
\vpointer &\displaystyle\frac{1}{2\lambda S_{w}(\theta) -\nu(\theta)}=\frac{r^2(\theta)}{\nu^2(\theta)} = \frac{(2\lambda S_{w}(\theta)+\eta'A(\theta)+\eta_0)^2}{\nu^2(\theta)}+\frac{(\eta'B(\theta))^2}{\nu^2(\theta)},\\
&\text{where $r^2(\theta) = (2\lambda S_{w}(\theta)+\eta'A(\theta)+\eta_0)^2+(\eta'B(\theta))^2$ in (\ref{rsquare.eq_01}).}\\
\end{array}
\end{equation}
\end{figure*}
}

\begin{figure*}
\begin{equation}
\begin{array}{ll}\label{max_condition}
(\lambda):& \displaystyle\frac{1}{2\pi}\int_{-\pi}^\pi S_w(\theta)\left(\frac{1}{2\lambda S_{w}(\theta)-\nu(\theta)}-2\frac{2\lambda S_{w}(\theta)+\eta'A(\theta)+\eta_0}{\nu(\theta)}+1\right)d\theta=P\\
(\eta): & -\displaystyle\frac{1}{2\pi}\int_{-\pi}^\pi\left(A(\theta)\frac{2\lambda S_{w}(\theta)+A'(\theta)\eta+\eta_0}{\nu(\theta)}+B(\theta)\frac{B'(\theta)\eta}{\nu(\theta)} \right)d\theta=0 \\
(\eta_0):& -\displaystyle\frac{1}{2\pi}\int_{-\pi}^\pi\frac{2\lambda S_{w}(\theta)+\eta'A(\theta)+\eta_0}{\nu(\theta)} d \theta+1=0\\
(\nu(\theta)):&\displaystyle\frac{1}{2\lambda S_{w}(\theta) -\nu(\theta)}=\frac{(2\lambda S_{w}(\theta)+\eta'A(\theta)+\eta_0)^2}{\nu^2(\theta)}+\frac{(\eta'B(\theta))^2}{\nu^2(\theta)} =\frac{r^2(\theta)}{\nu^2(\theta)}.
\end{array}
\end{equation}
\end{figure*}

\subsubsection{recover the optimal primal value}
$ $\\

Now, we are ready to show the strong duality. The approach is to find a dual solution $(\lambda^*, \eta^*,\eta^*_0, \nu^*(\theta))$ based on which we construct a primal solution $(x^*(\theta), y^*(\theta),c^*(\theta))$. Then we show the dual value and primal value coincide.
%

Assume $(\lambda^*, \eta^*,\eta^*_0, \nu^*(\theta))$ is an optimal solution of (\ref{Lagdual_b.eq}) (or (\ref{Lagdual.eq})). Then we construct
\begin{eqnarray}\label{xy_dual}
x^*(\theta)&=&\frac{2\lambda^* S_{w}(\theta)+\eta^{*'}A(\theta)+\eta^*_0}{\nu^*(\theta)},  \quad   a.e \label{xt.eq}\\
y^*(\theta)&=&\frac{\eta^{*'}B(\theta)}{\nu^*(\theta)},  \quad  a.e. \label{yt.eq}\\
c^*(\theta)&=&0,  \quad  a.e. \label{ct.eq}
\end{eqnarray}
Notice that $x^*(\theta),y^*(\theta), c^*(\theta)$ have the same form of (\ref{x.eq}), (\ref{y.eq}) and (\ref{c.eq}), respectively, and, with this association, we see that (\ref{max_condition}) coincide with the primal constraints of $(P)$. Moreover, from the fourth expression in (\ref{max_condition}), we obtain
\begin{equation}\label{nut.eq}
x^{*2}(\theta)+y^{*2}(\theta)=\frac{1}{2\lambda^* S_{w}(\theta)-\nu^*(\theta)}  \quad a.e.
\end{equation}
Since $\int_{-\pi}^\pi \log(2\lambda^* S_{w}(\theta)-\nu^*(\theta))d\theta $ is bounded, we have bounded $\int_{-\pi}^\pi 2\lambda^* S_{w}(\theta)-\nu^*(\theta)d\theta$ and thus bounded $\int_{-\pi}^\pi x^{*2}(\theta)+y^{*2}(\theta)d\theta$. This further indicates $x(\theta),y(\theta)\in \mathcal{L}_2$. Therefore $x^*(\theta), y^*(\theta)$  are feasible for (P). The primal cost achieved by this primal feasible solution is
\begin{equation*}
\begin{split}
&\frac{1}{2\pi}\int_{-\pi}^\pi -\frac{1}{2} \log(x^{*2}(\theta)+y^{*2}(\theta))d\theta\\
=&\frac{1}{2\pi}\int_{-\pi}^\pi \frac{1}{2}\log(2\lambda^* S_{w}(\theta)-\nu^*(\theta))d\theta.\\
\end{split}
\end{equation*}

{Next we show that
$$
\frac{1}{2\pi}\int_{-\pi}^\pi \frac{1}{2}\log(2\lambda^* S_{w}(\theta)-\nu^*(\theta))d\theta
$$
is equal to the optimal dual cost.

In particular, we use the expressions for $x^*(\theta)$ and $y^*(\theta)$ in (\ref{xt.eq}), (\ref{yt.eq}) and take substitutions to obtain (\ref{equ01_L})
\begin{figure*}
\begin{equation}
\begin{split}
L =&\frac{1}{2\pi}\int_{-\pi}^\pi\left[-\frac{1}{2}\log\left(x^{*2}(\theta)+y^{*2}(\theta)\right)+\frac{1}{2}-\frac{(2\lambda^* S_{w}(\theta)+\eta^{*'}A(\theta)+\eta^*_0)^2 }{2\nu^*(\theta)}-\frac{(\eta^{*'}B(\theta))^2}{2\nu^*(\theta)} +\lambda^*  S_{w}(\theta)\right]d\theta-\lambda^* P+\eta^*_0.\\
\end{split}
\label{equ01_L}
\end{equation}
\end{figure*}
with constraints under the maximum conditions
\begin{equation}
\begin{array}{ll}
&\frac{1}{2\pi}\int_{-\pi}^\pi S_{w}(\theta)\left(x^{*2}(\theta)+y^{*2}(\theta)-2x^*(\theta)+1\right)d\theta=P,\\
& \frac{1}{2\pi}\int_{-\pi}^\pi(A(\theta)x^*(\theta)+B(\theta)y^*(\theta))=0,\\
&\frac{1}{2\pi}\int_{-\pi}^\pi x^*(\theta)d\theta=1.\\
 \end{array}
\label{max_condition_solution}
\end{equation}
That is,
$$
L=\frac{1}{2\pi}\int_{-\pi}^\pi-\frac{1}{2}\log(x^{*2}(\theta)+y^{*2}(\theta))d\theta+X,
$$
where $X$ is defined in (\ref{def_X}).
\begin{figure*}
\begin{equation}
\begin{split}
X=&\frac{1}{2\pi}\int_{-\pi}^\pi\left[\frac{1}{2}-\frac{(2\lambda^* S_{w}(\theta)+\eta^{*'}A(\theta)+\eta^*_0)^2 }{2\nu(\theta)}-\frac{(\eta{*'}B(\theta))^2}{2\nu^*(\theta)}+\lambda^*  S_{w}(\theta)\right]d \theta-\lambda^* P+\eta^*_0\\
=& \frac{1}{2\pi}\int_{-\pi}^\pi\left[\frac{1}{2}-\frac{(2\lambda^* S_{w}(\theta)+\eta^{*'}A(\theta)+\eta^*_0) }{2}x^{*}(\theta)-\frac{(\eta{*'}B(\theta))}{2}y^{*}(\theta)+\lambda^*  S_{w}(\theta)\right]d \theta-\lambda^* P+\eta^*_0.\\
\end{split}
\label{def_X}
\end{equation}
\end{figure*}
Next we show that $X=0$. From equations (\ref{reducex.eq}), (\ref{reducey.eq}) and (\ref{max_condition_solution}),
$$
\begin{array}{ll}
X=&\frac{1}{2\pi}\int_{-\pi}^\pi\left(\frac{1}{2}+\frac{\nu^*(\theta)}{2}(x^{*2}(\theta)+y^{*2}(\theta)) \right)d\theta +\lambda^*\left[\frac{1}{2\pi}\int_{-\pi}^\pi S_{w}(\theta) (1-2 x^*(\theta))d\theta - P\right].
 \end{array}
$$
Adding and subtracting $\lambda S_w(\theta)(x^{*2}(\theta)+y^{*2}(\theta))$ we have (\ref{equ:strong_dual_01}).
\begin{equation}
\begin{split}
X=&\frac{1}{2\pi}\int_{-\pi}^\pi\left(\frac{1}{2}+\frac{\nu^*(\theta)}{2}(x^{*2}(\theta)+y^{*2}(\theta))-\lambda^* S_{w}(\theta)(x^{*2}(\theta)+y^{*2}(\theta))\right)d\theta\\
& +\lambda^*\left[\frac{1}{2\pi}\int_{-\pi}^\pi S_{w}(\theta)\left(x^{*2}(\theta)+y^{*2}(\theta) -2x^*(\theta)+1\right)d\theta - P\right].\\
\end{split}
\label{equ:strong_dual_01}
\end{equation}
From equations (\ref{rsquare.eq}), (\ref{rsquare.eq_ext}) and (\ref{max_condition_solution}), one can check that $X=0$.
Therefore the  dual cost is equal to the primal cost.

\subsection{Proof of Lemma \ref{lemma.exist}}
\label{proof:existence}
Recall that the dual problem (\ref{Lagdual_c.eq}) for the primal problem (\ref{proof:strong_dual_primal_problem}) is given by,
\begin{equation}
\begin{array}{ll}
\mu_h^D &= \sup_{\lambda, \eta, \eta_0} \inf_{x,y} L(x, y, c, \lambda, \eta, \eta_0)\\
&=\sup_{\lambda, \eta, \eta_0}g(\lambda,\eta,\eta_0),\\
\end{array}
\end{equation}
where $\lambda > 0, \eta = [\eta_1,\eta_2,\cdots, \eta_h]' \in \mathbb{R}^h, \eta_0 \in \mathbb{R}$ and the Lagrangian function $L$ is given by
(\ref{lap.eq}) with $c(\theta)=0$.

Clearly, $\mu_h^D$ is a lower bound on the primal problem (\ref{proof:strong_dual_primal_problem}). Also, it can be verified that the optimal $\mu_h^D > -\infty$ by plugging a specific $(\lambda, \eta, \eta_0)$ into $g(\lambda,\eta,\eta_0)$, e.g., $\lambda = 1, \eta = [0,0,\cdots, 0]' \in \mathbb{R}^h, \eta_0 = 0$. Next, we show by contradiction that the optimal dual variables are bounded (or exist).

First of all, recall a filter defined by
$$\mathbb{Q}(e^{j\theta}) = a(\theta) + j b(\theta) = \sum_{n=-\infty}^{+\infty} c_n e^{-j\theta}.$$
Let $x(\theta) = a(\theta) +1, y(\theta)=b(\theta)$. From (\ref{proof:strong_dual_primal_problem}) we know the equality constraints
$$\frac{1}{2\pi}\int_{-\pi}^\pi(A(\theta)x(\theta)+B(\theta)y(\theta))d\theta = 0, \quad \frac{1}{2\pi}\int_{-\pi}^\pi x(\theta) d\theta =1$$
impose zeros on the Fourier coefficients $c_0, c_{-1}, c_{-2},\cdots, c_{-h}$ of the filter $Q(e^{j\theta})$. Specifically,
$$c_0 = \frac{1}{2\pi}\int_{-\pi}^\pi x(\theta) d\theta -1,$$
$$c_{-n} = \frac{1}{2\pi}\int_{-\pi}^\pi(\cos(n\theta)x(\theta)+\sin(n\theta)y(\theta))d\theta, $$ for $n= 1,2,\cdots,h$.
Also, the constraint $$\tilde{P}(x,y):= \frac{1}{2\pi}\int_{-\pi}^\pi S_{w}(\theta)(x^2(\theta)+y^2(\theta)-2x(\theta)+1)d\theta - P$$
imposes the power cost of the filter less than $P$. Equivalently, the Lagrangian function (\ref{lap.eq}) can be rewritten as
\begin{equation}\label{equ:L_cn}
\begin{array}{ll}
L =&  \displaystyle\frac{1}{2\pi}\int_{-\pi}^\pi -\frac{1}{2}\log(x^2(\theta)+y^2(\theta))d\theta + \lambda \tilde{P}(x,y) + \sum_{n=1}^{h} \eta_n c_{-n} + \eta_0 c_0.\\
\end{array}
\end{equation}

Now, assume at least one of the optimal dual variables is unbounded, i.e., there exists a sequence $\xi_m = (\lambda_m, \eta_m, \eta_{0,m})$ with $\lim_{m \rightarrow \infty} ||\xi_m||_2 = \infty$ that satisfies
\begin{equation*}
\begin{split}
\mu_h^D &= \lim_{m \rightarrow \infty} g(\xi_m)  =  \lim_{m \rightarrow \infty} \min_{x,y} L(x, y, \xi_m) > -\infty.\\
\end{split}
\end{equation*}

It is worth noting that the case of unbounded $\lambda$ can be trivially ruled out since $ \tilde{P}(x,y)$ can be always negative with a feasible $(x,y)$. Next, for each $\xi_m$, we now construct a filter $Q_m(e^{j\theta}) = \tilde{x}_m(\theta) - 1 + j \tilde{y}_m(\theta)$ by assigning a sequence of Fourier coefficients $c_{n}$ for $n\in \mathbb{Z}$ where $c_{-n}$ for $n = 0,1,\cdots, h$ has an \textit{opposite} sign to its corresponding dual variable as shown in (\ref{equ:L_cn}). This implies the term $\sum_{n=1}^{h} \eta_n c_{-n} + \eta_0 c_0 \rightarrow -\infty$ as $m$ increases. Note that, since the dual variable $\lambda > 0$, we construct the filter satisfying the power constraint by scaling down $c_n$ if needed. Therefore, if $\lim_{m \rightarrow \infty} ||\xi_m||_2 = \infty$, we have
\begin{equation*}
\begin{split}
\mu_h^D &=  \lim_{m \rightarrow \infty} \min_{x,y} L(x, y, \xi_m) \leq  \lim_{m \rightarrow \infty} L(\tilde{x}_m, \tilde{y}_m, \xi_m) = -\infty. \\
\end{split}
\end{equation*}
This contradicts $\mu_h^D > -\infty$. The proof is complete.

\subsection{Proof of Corollary \ref{coro:upper_bound}}
From the solution of (\ref{opt_upperbound_approximate}),  $\lambda ^{(m)}, \eta^{(m)}, \eta_0^{(m)}, \nu_i^{(m)}$, we know $\lambda ^{(m)}, \eta^{(m)}, \eta_0^{(m)}$ are feasible  for (D) in (\ref{dual.eq}), and any feasible dual solution provides a cost $- g(\lambda ^{(m)}, \eta^{(m)}, \eta_0^{(m)})$, which is an upper bound on $C_{fb}(h)$. Namely, $\overline{C_{fb}(m,h)} \geq C_{fb}(h)$.

Next, consider a sequence $\lbrace \lambda ^{(m)}, \eta^{(m)}, \eta_0^{(m)}, \nu_i^{(m)} \rbrace_{m=1}^{\infty}$ of optimal solutions of  (\ref{opt_upperbound_approximate}) for each $m$. Then, the fact that $\lim_{m\rightarrow \infty} \mu_h(m) = \mu_h$ implies that $\lim_{m\rightarrow \infty} \max_{\nu(\theta)\geq 0}\tilde{g}(\lambda ^{(m)}, \eta^{(m)}, \eta_0^{(m)}, \nu(\theta)) = -\mu_h$. Based on the equivalence of $a)$ and $b)$ in Theorem \ref{strongdual.thm}, we further have
$\lim_{m\rightarrow \infty} g(\lambda ^{(m)}, \eta^{(m)}, \eta_0^{(m)}) = -\mu_h$.

Therefore, we have $\lim_{m\rightarrow\infty}\overline{C_{fb}(m,h)} = \mu_h$. According to $c)$ in Theorem \ref{strongdual.thm}, i.e., $C_{fb}(h) = \mu_h$, and Lemma \ref{lemma:C_h}, the result directly follows.

\subsection{Proof of Theorem \ref{thm:lower_bound_convergence}} \label{appendix: proof_LB}

{ The main idea of the proof is the following: we first approximate the feedback capacity characterization (\ref{capacity_long01}), $C_{fb}$,  by invoking a FIR filter $\mathbb{Q}$ of order $N$ and sampling the filter in frequency domain with equal spacing $\frac{\pi}{m}$. We denote this two-degree approximation by $C_{fb}^N(m)$ and prove that $C_{fb}^N(m)$ converges to $C_{fb}$ as $N$ and $m$ increase. Then we show the constructed code is essentially an optimal solution to $C_{fb}^N(m)$, implying that the proposed feedback code is $C_{fb}$-achieving as $N$ and $m$ are sufficiently large. We next prove Theorem \ref{thm:lower_bound_convergence} by providing detailed derivations in each aforementioned step.}\\

\subsubsection{A Primal Optimization Problem Equivalent to $\mu_h(m)$}
$ $\\

First of all, we present a finite dimensional optimization and we show that the dual of the optimization is equivalent to (\ref{opt_upperbound_approximate}). This is a necessary result in order to show that the proposed code construction is essentially an optimal solution to $C_{fb}^N(m)$

We start from (\ref{capacity_long01}), rewritten as follows

\begin{equation}
\begin{split}
C_{fb}=&\max_{\mathbb{U},\mathbb{Q}}\frac{1}{2\pi}\int_{-\pi}^{\pi}\frac{1}{2}\log\left( \mathbb{U}(e^{j\theta})+|1+\mathbb{Q}(e^{j\theta})|^2\right)d\theta,\\
s.t. \quad  &\frac{1}{2\pi}\int_{-\pi}^{\pi}\left(\mathbb{U}(e^{j\theta})+|\mathbb{Q}(e^{j\theta})|^2\right)\mathbb{S}_w(e^{j\theta})d\theta\leq P,\\
& \mathbb{U}(e^{j\theta})\geq 0,\;\mathbb{Q}(e^{j\theta})\in \mathcal{RH}_2,\, \text{strictly causal},\\
\end{split}
\label{capacity_long03}
\end{equation}
where $\mathbb{U}(e^{j\theta})=\frac{\mathbb{S}_v(e^{j\theta})}{\mathbb{S}_w(e^{j\theta})}$ since by assumption $\mathbb{S}_w(e^{j\theta})>0$ for all $\theta\in [-\pi,\pi]$.
Letting $c^2(\theta)=\frac{\mathbb{S}_v(e^{j\theta})}{\mathbb{S}_w(e^{j\theta})}$ and following the same steps as in (\ref{formula_stationaryGuassian_upperbound_equi}), we obtain its generalization as
\begin{equation}
\begin{split}
C_{fb}=&\max_{\Gamma}\frac{1}{4\pi}\int_{-\pi}^{\pi}\log (c^2(\theta)+(1+a(\theta))^2+ b(\theta)^2 )d\theta\\
s.t. \quad &\frac{1}{2\pi}\int_{-\pi}^{\pi}\left(a^2(\theta)+b^2(\theta)+c^2(\theta)\right) S_w(\theta)d\theta\leq P,\\
&\int_{-\pi}^{\pi} a(\theta)\cos(n\theta) d\theta + \int_{-\pi}^{\pi} b(\theta) \sin(n\theta)d\theta = 0,\quad n\geq 0\\
\Gamma =& \lbrace a(\theta), b(\theta), c(\theta): [-\pi, \pi] \rightarrow \mathbb{R} |  a(\theta), b(\theta), c(\theta)\in \mathcal{L}_2\rbrace.
\end{split}
\label{new.eq}
\end{equation}
Note that we recover (\ref{formula_stationaryGuassian_upperbound_equi}) if $c(\theta)=0$. We are now ready to consider a finite-dimensional approximation of the above problem. Define
\begin{equation}\label{cfbhfd.eq}
\begin{split}
C_{fb}(m,h)=&\max_{a_i,b_i,c_i\in \mathbb{R}}\frac{1}{4m}\sum_{i=1}^{2m}\log ((1+a_i)^2+ b_i^2+c_i^2 )\\
s.t. \quad &\frac{1}{2m}\sum_{i=1}^{2m}\left(a^2_i+b^2_i+c_i^2\right) S_w(\theta_i) \leq P,\\
&\frac{1}{2m}\sum_{i=1}^{2m}a_i\cos(n\theta_i) +\frac{1}{2m} \sum_{i=1}^{2m} b_i \sin(n\theta_i) = 0, \\
& \quad n = 0, 1,2,\cdots, h. \quad  \theta_i = -\pi+ \frac{\pi}{m}(i-1).
\end{split}
\end{equation}
As a result of the power constraint, the feasible set of $\lbrace a_i,b_i,c_i\in \mathbb{R} \rbrace$ is compact. Therefore, the existence of an optimal solution is guaranteed.
\begin{lemma}\label{lemma_LMI}
(\ref{cfbhfd.eq}) is equivalent to the following convex optimization problem:
 \begin{equation}\label{cfbhfd5.eq}
\begin{split}
C_{fb}(m,h)=&\max_{W_{i},x_i,b_i\in \mathbb{R}}\frac{1}{4m}\sum_{i=1}^{2m}\log (W_{i})\\
s.t. \quad &\frac{1}{2m}\sum_{i=1}^{2m}\left(W_{i}-2x_i+1\right) S_w(\theta_i) \leq P,\\
&\frac{1}{2m}\sum_{i=1}^{2m}x_i=1\\
&\frac{1}{2m}\sum_{i=1}^{2m}x_i\cos(n\theta_i) + \frac{1}{2m}\sum_{i=1}^{2m} b_i \sin(n\theta_i) = 0\\
& \quad n = 1,2,\cdots, h.\\
&\theta_i = -\pi+ \frac{\pi}{m}(i-1)\\
&\left[\begin{array}{ccc} W_{i} & x_i & b_i\cr x_i & 1 &0\cr b_i & 0 & 1\end{array} \right]\geq 0.
\end{split}
\end{equation}
\end{lemma}
\begin{IEEEproof}
Let $x_i=1+a_i$ and $W_i=x_i^2+b_i^2+c_i^2$. Since $c_i^2$ is free, $W_i$ must equivalently satisfy
$W_i\geq x_i^2+b_i^2$, which can be rewritten as an LMI by the Schur complement. The feasible sets of the two problems are equivalent by simple transformations.
\end{IEEEproof}
Problem (\ref{cfbhfd5.eq}) can be solved efficiently using standard interior point methods. It is interesting to note that if the optimal solution satisfies the Linear Matrix Inequality constraint with equality for all $i=1,\ldots, 2m$, then the solution exclusively uses feedback, as the $c_i$'s, which represent the feedforward component, will be all equal to zero. On the other hand, it is possible that the optimal solution of these finite-dimensional optimization  may require $c_i\neq 0$ for some $i$. { More discussion can be found in the proof of Corollary \ref{Cor}} \label{coro.discrete.dual.val}.}

\begin{lemma}\label{lemma:lower_convergence_equivalent_opt01}
Given $m>0$ and $h>0$, $\mu_h(m)$ in (\ref{opt_upperbound_approximate}) is a dual optimization of $C_{fb}(m,h)$ in (\ref{cfbhfd.eq}), and $\mu_h(m)=C_{fb}(m,h)$.\\
\end{lemma}

\begin{IEEEproof}
See Appendix \ref{lem.strong.dual}.
\end{IEEEproof}
\begin{corollary}\label{Cor}
Given $h$ and $m$, let $a_i$, $b_i$ and $c_i$, $i=1,\ldots 2m$ be the solution to (\ref{cfbhfd.eq}), obtained from $W_i$, $x_i$, $b_i$, the solution of (\ref{cfbhfd5.eq}) with $a_i=x_i-1$, and $c_i=W_i-x_i^2-b_i^2$.
Let $\lambda, \eta, \eta_0$, and $\nu_i$, $i=1,\ldots 2m$,  be the optimal solution to $\mu_h(m)$ in (\ref{opt_upperbound_approximate}).
Then for all $i$'s such that $\nu_i>0$,
\begin{equation}\label{thm_xy_dual_approx_2}
\begin{split}
a_i&=\frac{2\lambda {S}_{w}(\theta_i)+\eta'A(\theta_i)+\eta_0}{\nu_i}-1, \\
b_i&=\frac{\eta'B(\theta_i)}{\nu_i}.
\end{split}
\end{equation}
\end{corollary}
\begin{IEEEproof}
See Appendix \ref{coro.discrete.dual.val}.
\end{IEEEproof}

\begin{remark}
The reader should notice the similarity with (\ref{thm_xy_dual})  obtained in the semi-infinite dimensional case.  Note however, that the primal solution to $C_{fb}(m,h)$ may require contribution from both feedforward and feedback components while the solution to $C_{fb}(h)$ is always guaranteed to be exclusively obtained through feedback when the channel is not flat. Based on Lemma \ref{lemma:lower_convergence_equivalent_opt01}, therefore, for fixed $h$ and large enough $m$ the solution to $C_{fb}(m,h)$ will tend to be exclusively generated from feedback.\\
\end{remark}

\subsubsection{FIR approximations on $C_{fb}$}
$ $

Since FIR solutions are dense in ${\cal RH}_2$,
$C_{fb}$ in (\ref{capacity_long03}) can be arbitrarily (uniformly) well approximated by an FIR $\mathbb{Q}$ of order $N$, for $N$  large enough. This motivates us to introduce
\begin{equation}
\begin{split}
C^N_{fb}=&{\max}_{\mathbb{U}_N,\mathbb{Q}_N}\frac{1}{2\pi}\int_{-\pi}^{\pi}\frac{1}{2}\log\left( \mathbb{U}_N(e^{j\theta})+|1+\mathbb{Q}_N(e^{j\theta})|^2\right)d\theta,\\
s.t. \quad  &\frac{1}{2\pi}\int_{-\pi}^{\pi}\left(\mathbb{U}_N(e^{j\theta})+|\mathbb{Q}_N(e^{j\theta})|^2\right)\mathbb{S}_w(e^{j\theta})d\theta\leq P,\\
& \mathbb{U}_N(e^{j\theta})\geq 0, \\
&{ \mathbb{Q}_N(e^{j\theta})= \displaystyle\sum_{k=1}^N q_k e^{-kj\theta};\;\text{FIR order N.}}
\end{split}
\label{capacity_longFIR_N}
\end{equation}

\begin{remark}
An optimal solution to the above problem exists since the solution of (\ref{capacity_long01}) exists and the set of FIR of order $N$ is closed in ${\cal H}_2$.
\end{remark}

Therefore,  we have
\begin{equation}
\lim_{N \to \infty}C_{fb}^N=C_{fb}.
\label{FIR_approx_converge}
\end{equation}
Now, from the solution to (\ref{capacity_longFIR_N}) we define
$$
\tilde{C}_{fb}^N=\frac{1}{2\pi}\int_{-\pi}^\pi\log |1+\mathbb{Q}_N(e^{j\theta})|d\theta.
$$
Note the above integral is well-defined as ${\mathbb{Q}}_N$ is FIR. Clearly, we have $C_{fb}\geq C_{fb}^N\geq \tilde{C}_{fb}^N$.
Moreover, from (\ref{capacity_short01}) and Remark \ref{sv0.rem},  it follows that for $N$ large enough the contribution to the capacity of the feedforward component $\mathbb{U}_N$ must be zero or going to zero (with arbitrarily small power allocation).
From the above property,  Remark \ref{rem:unit_zero} and the fact that the  Fourier series uniformly converge to $\mathbb{RL}_2$ functions, we expect $1+\mathbb{Q}_N$ to have no zeros on the unit circle for large enough $N$.
For simplicity we make this a standing assumption. Therefore, we have
\begin{equation}
\lim_{N\rightarrow \infty} \tilde{C}_{fb}^N -C_{fb}^N = 0.
\label{FIR_approx_nullForward_converge}
\end{equation}

\subsubsection{Finite-dimensional approximations on $C_{fb}$}
$ $\\
{ Firstly, we note that the discretize-then-optimize methodology has been widely investigated and used to solve optimization problems involving partial differential equations (PDEs) with integration. An overview can be found in \cite{Ghobadi_2014,PDE_liu2017} and the reference therein. In what follows, we utilize this methodology in our proof. In particular,} a finite dimensional approximation of $C_{fb}$ in (\ref{capacity_long03}) is obtained for large $N$ and by fine frequency discretization of (\ref{capacity_longFIR_N}),
\begin{equation}\label{primalFIR.eq}
\begin{split}
C_{fb}^N(m)=&\max_{\mathbb{U}^m_N(e^{j\theta_i})\geq 0, q_k^m}\frac{1}{2m}\sum_{i=1}^{2m}\frac{1}{2}\log\left(\mathbb{U}^m_N(e^{j\theta_i})+ |1+ \mathbb{Q}_N^m(e^{j\theta_i})|^2\right)\\
s.t. \quad &\frac{1}{2m}\sum_{i=1}^{2m}\left((\mathbb{U}^m_N(e^{j\theta_i})+|\mathbb{Q}_N^m(e^{j\theta_i})|^2\right) \mathbb{S}_w(e^{j\theta_i}) \leq P,\\
& \mathbb{Q}_N^m(e^{j\theta_i}) = \displaystyle\sum_{k=1}^N q_k^m e^{-kj\theta_i}, \quad q_k^m\in \mathbb{R},\\
&\theta_i = -\pi+ \frac{\pi}{m}(i-1).\\
\end{split}
\end{equation}
It follows that\footnote{
Assume the contrary, i.e., there is always an $m$ for which
\begin{equation}\label{contra1.eq}
\begin{array}{l}
\mbox{ either }  C_{fb}^N - C^N_{fb}(m)\geq \gamma,\\
\mbox{ or }  C_{fb}^N(m) - C^N_{fb}\geq \gamma
\end{array}
\end{equation}
for some fixed $\gamma>0$.  Take the optimal solution to (\ref{capacity_longFIR_N}). Since $\mathbb{Q}_N(e^{j\theta})$ is continuous and bounded, $1+\mathbb{Q}_N$ has no zeros on the unit circle, and $\mathbb{U}_N(e^{j\theta})\mathbb{S}_w(e^{j\theta})$, (if nonzero), water-fills the modified noise spectrum
$|1+\mathbb{Q}_N(e^{j\theta})|^2\mathbb{S}_w(e^{j\theta})$.
Both integrals in (\ref{capacity_longFIR_N}) can be arbitrarily well approximated by the corresponding sums since the integrands are Riemann integrable. But this means that within a small perturbation, (diminishing to zero  with $m$ increasing to infinity) we can always find, for all  $m$ large enough, a feasible solution to (\ref{primalFIR.eq}) with cost within any $\epsilon>0$, from  $C_{fb}^N$, a fact that contradicts the first inequality in  (\ref{contra1.eq}). To invalidate the second inequality, take the sequence of such $m$'s  and call it $m_s$. Assume $\{q_k^{m_s}\}_{k=1}^N$ leads to $1+\mathbb{Q}_N^{m_s}$ with no zeros on the unit circle (otherwise perturb it).  $\mathbb{U}_N^{m_s}(e^{j\theta_i})\mathbb{S}_w(e^{j\theta_i})$ will water-fill
$|1+\mathbb{Q}_N^{m_s}(e^{j\theta_i})|^2\mathbb{S}_w(e^{j\theta_i})$
Now take any $\delta>0$ and an large enough $m_s$,  and consider $\mathbb{Q}_N^{m_s}(e^{j\theta})$ and the corresponding water-filling solution $\mathbb{U}_N^{m_s}(e^{j\theta})$ to a budget power $P+\delta$. This is always possible. Now since the Riemann sums are arbitrarily close to the integrals, the cost will be arbitrarily close to $C_{fb}^N+\gamma$. But this is a contradiction, as $\delta$ was arbitrary, and we know that with power $P$ the largest rate cannot be strictly greater than  $C_{fb}^N$.  }
\begin{equation}\label{FIR_approx_converge.eq01}
\lim_{m\to \infty} C_{fb}^N(m) = C^N_{fb},
\end{equation}
According to (\ref{FIR_approx_converge}), we conclude that
$$\lim_{N\to \infty}\lim_{m \to \infty}C_{fb}^N(m)= C_{fb}.$$
Similarly to $\tilde{C}_{fb}$, we introduce\footnote{We assume $1+\mathbb{Q}_N^m(e^{j\theta_i})\neq 0$  for all $i$. Otherwise, one can apply an arbitrarily small perturbation to $\{q_k^m\}_{k=1}^N$ (resulting in a nearly optimal solution) to exclude unit-circle zeros $\mathbb{Q}_N^m(e^{j\theta})$.
In fact, as a consequence of Remark \ref{rem:unit_zero}, this assumption always holds for sufficiently large $m,N$}
\begin{equation}\label{primalFIR.eq_short}
\tilde{C}_{fb}^N(m)=\frac{1}{2m}\sum_{i=1}^{2m}\log|1+ \mathbb{Q}_N^m(e^{j\theta_i})|.\\
\end{equation}

Note that $\tilde{C}^N_{fb}(m)\leq C_{fb}^N(m)$. We next show that
\begin{equation}\label{finite_approx_converge.eq01}
\lim_{N\to \infty}\lim_{m \to \infty}\tilde{C}_{fb}^N(m)= C_{fb}.
\end{equation}
We present the proof as follows. Since we know that $\lim_{N\to \infty}\lim_{m \to \infty}C_{fb}^N(m)= C_{fb}$, we have
$$\limsup_{N\to \infty}\limsup_{m \to \infty}\tilde{C}_{fb}^N(m)\leq  C_{fb}.$$
We claim that equality holds. Assume by contradiction that we can always find $N$ and $m>>N$ such that
$$
C_{fb}-\tilde{C}_{fb}^N(m)\geq \gamma>0,
$$
while $|C_{fb}-C_{fb}^N(m)|\leq \epsilon$.
$C_{fb}-\tilde{C}_{fb}^N(m)\geq \gamma$ implies that
$$
P_{fN}^m=\frac{1}{2m}\sum_{i=1}^{2m}\mathbb{U}^m_N(e^{j\theta_i}) \mathbb{S}_w(e^{j\theta_i})\geq \eta>0.
$$
However, omitting some of the tedious steps,
which follow similar previous arguments, this would imply that nonzero feedforward power is required to achieve capacity. Definitely this contradicts the fact that, under our assumption of non-white noise spectrum, such power allocation cannot be capacity achieving.

Putting (\ref{FIR_approx_converge}), (\ref{FIR_approx_converge.eq01}) and (\ref{finite_approx_converge.eq01}) together, we have that, for a given $\epsilon>0$, there exist large enough $N$ and $m$ such that
\begin{equation}\label{inequ:3epsilon}
\begin{split}
&|C_{fb}-\tilde{C}_{fb}^N(m)| =|C_{fb} - C_{fb}^N + C_{fb}^N - C_{fb}^N(m) + C_{fb}^N(m) -\tilde{C}_{fb}^N(m)|<3\epsilon.\\
\end{split}
\end{equation}
$ $\\

\subsubsection{Proof of Theorem \ref{thm:lower_bound_convergence}}
$ $\\

First of all, we need the following lemma to prove our theorem.
\begin{lemma}\label{lemma:lower_convergence_equivalent_opt}
Given $N\in \mathbb{Z}_+$ with $N<2m$, the optimization $C_{fb}^N(m)$ in (\ref{primalFIR.eq}) is equivalent to the optimization $C_{fb}(m,h)$ in (\ref{cfbhfd.eq})  with $h=2m-N-1$, i.e.
$$
C_{fb}^N(m)=C_{fb}(m,2m-N-1).
$$
\end{lemma}

\begin{IEEEproof}
We start with (\ref{primalFIR.eq}). Let $\mathbb{Q}_N^m(e^{j\theta_i}) = a_i+jb_i$ and $c_i^2 = \mathbb{U}^m_N(e^{j\theta_i})$. { The proof follows from the properties of the DFT applied to the periodic spectrum (using the synthesis and analysis equations in Section 4.21 \cite{proakis4thEdition}). Specifically, we follow the proof of Lemma \ref{lemma_symmetric_filter} (in particular, follow the discretized (\ref{equ.coeff.equality01})) by imposing zeros on Fourier coefficients $q_k^m = \displaystyle\frac{1}{2m}\sum_{i=1}^{2m}a_i\cos(k\theta_i)-b_i\sin(k\theta_i)$ for $k=N+1, \cdots, 2m$.}
\end{IEEEproof}

In what follows, we follow the three-step procedure of feedback codes construction. Based on Lemma \ref{lemma:lower_convergence_equivalent_opt01} and Lemma \ref{lemma:lower_convergence_equivalent_opt}, we have
$$C_{fb}^N(m)=C_{fb}(m,2m-N-1)= \mu_{2m-N-1}(m).$$

This implies that the feedback code constructed from the optimal solution to $\mu_{2m-N-1}(m)$ is an optimal solution to $C_{fb}^N(m)$. In particular, let $a,b$ be the optimal solution of $C_{fb}^N(m)$. $a,b$ can be computed from (\ref{cfbhfd.eq}), or directly from (\ref{thm_xy_dual_approx}) for $\nu_{i,h,m}>0$, as described in the first step of feedback codes construction.

Then, $\mathbb{Q}^m_N(e^{j\theta})$  can be computed from  (\ref{alg:L2_filter}) as described in the third step, and requires power
$$
p^m_N=\frac{1}{2\pi}\int_{-\pi}^\pi |\mathbb{Q}^m_N(e^{j\theta})|^2\mathbb{S}_w(e^{j\theta}) d\theta.
$$
In addition, $\mathbb{Q}^m_N(e^{j\theta_i})$ will satisfy the power constraint of (\ref{primalFIR.eq}) and thus,
$$
\frac{1}{2m}\sum_{i=1}^{2m}|\mathbb{Q}_N^m(e^{j\theta_i})|^2 \mathbb{S}_w(e^{j\theta_i}) = P-P_{fN}^m,
$$
since there may exist few $c_i=\mathbb{U}_N^m(e^{j\theta_i})\geq 0$.

Then,  $\eta^m_N = p^m_N-(P-P_{fN}^m)$ must go to zero with $m$ given that $N$ is fixed, as the spectrum of an FIR of order $N$ can be arbitrarily well approximated by $m$ samples with $m>>N$. In addition, the rate for such $\mathbb{Q}^m_N(e^{j\theta})$ is given by
$$
\hat{C}_{fb}^N(m)=\frac{1}{2\pi}\int_{-\pi}^{\pi}\log |1+\mathbb{Q}_N^m(e^{j\theta})|d\theta,
$$
and for $m>m_2$,
$$
|\hat{C}_{fb}^N(m)-\tilde{C}_{fb}^N(m)|<\epsilon.
$$
Combined with (\ref{inequ:3epsilon}), it yields that for $m>\max\{m_0,m_1,m_2\}$,
$$
|C_{fb}-\hat{C}_{fb}^N(m)|<4\epsilon.
$$

Now let $\delta^m_N= p^m_N-P$. If $\delta^m_N\leq 0$ then $\hat{C}_{fb}^N(m)$ is an achievable rate based on Theorem  \ref{thm_capacity_achieving_code}, within $4\epsilon$ from $C_{fb}$.
If $\delta^m_N>0$ then the power constraint is violated, albeit by a negligible amount for large enough $N$ and $m$.

We can rescale $\mathbb{Q}^m_N(e^{j\theta})$ by $\alpha_{N,m}=\sqrt{\frac{P}{p^m_N}} = \sqrt{1-\frac{\delta^m}{P+\delta^m_N}}$ so that $\alpha_{N,m}\mathbb{Q}^m_N(e^{j\theta})$ uses power $P$. When $\delta_N^m\leq 0$, $\alpha_{N,m}\geq 1$ and we improve the lower bound, $\hat{C}_{fb}^N(m)$, by the rescaling. When   $\delta_N^m> 0$, $\hat{C}_{fb}^N(m)$ is not a lower bound on the rate, however, $\alpha_{N,m}<1$,
and we obtain a lower bound using $\alpha_{N,m}\mathbb{Q}^m_N(e^{j\theta})$ in place of $\mathbb{Q}^m_N(e^{j\theta})$ .

Based on Theorem \ref{thm_capacity_achieving_code}, the last step of feedback code construction leads to a rate
$$
R_N(m)=\frac{1}{2\pi}\int_{-\pi}^\pi\log |1+\alpha_{N,m}\mathbb{Q}_N^m(e^{j\theta})|d\theta.
$$
We next note that
$$
\delta^m_N=p^m_N-P=p_N^m-P + P_{fN}^m - P_{fN}^m=\eta^m_N-P_{fN}^m.
$$
Furthermore,
$$
\delta^m_N=\eta^m_N-P_{fN}^m + P_{fN} - P_{fN},
$$
$$
|\delta^m_N|\leq|\eta^m_N|+|P_{fN}|+|P_{fN}^m-P_{fN}|,
$$
where $P_{fN} = \int_{-\pi}^{\pi} \mathbb{U}_N(e^{j\theta})\mathbb{S}_w(e^{j\theta})d\theta$.
As $m\to\infty$, $|\eta^m_N|\to 0$ and $|P_{fN}^m-P_{fN}|\to 0$.
Based on Remark \ref{sv0.rem}, as $N\to \infty$, $|P_{fN}|\to 0$. Thus
$$\lim_{N\to \infty}\lim_{m\to\infty}\delta_N^m =0,\mbox{ and } \lim_{N\to \infty}\lim_{m\to\infty}\alpha_{N,m} =1.$$

Therefore, for large enough $N$ and $m$,
$$
|\hat{C}_{fb}^N(m)-R_N(m)|<\epsilon,
$$
and
$$
C_{fb}-R_N(m)<5\epsilon.
$$

}
The proof is complete.

\subsection{Proof of Lemma \ref{lemma:lower_convergence_equivalent_opt01}} \label{lem.strong.dual}

We consider the dual of (\ref{cfbhfd5.eq}). Since the Slater's condition holds and we can always have a feasible primal solution, e.g.,$W_i = x_i=1$ and $b_i=0$, the dual problem has no gap. Moreover, the optimal dual solution exists.

From standard derivations of dual problems involving LMI constraints \cite{Boyd_book}, the dual problem is given by the following optimization,
\begin{equation}\label{cfbhfd5_dual.eq}
\begin{split}
C_{fb}(m,h)=-&\max_{\lambda\geq 0, \eta_0,\eta\in \mathbb{R}^h, \tilde{\nu}_i,\xi_{4i},\xi_{6i},\xi_{5i}}g^D(\lambda,\eta_0,\eta,\nu_i,\xi_{4i},\xi_{6i})\\
s.t. \quad &\Xi_i\geq 0,\; i=1,\ldots,2m,\\
&r_{1i}=2\lambda S_w(\theta_i)+\eta' A(\theta_i)+\eta_0,\\
&r_{2i}=\eta' B(\theta_i),\\
&\theta_i = -\pi+ \frac{\pi}{m}(i-1),
\end{split}
\end{equation}
where
\begin{equation}\label{xi.eq}
\Xi_i=\left[\begin{array}{ccc} \tilde{\nu}_{i} & -\frac{r_{1i}}{2} & -\frac{r_{2i}}{2}\cr -\frac{r_{1i}}{2} & \xi_{4i} &\xi_{5i}\cr -\frac{r_{2i}}{2} & \xi_{5i} & \xi_{6i}\end{array} \right],
\end{equation}
$$\begin{array}{l}
g^D(\lambda,\eta_0,\eta,\tilde{\nu}_i,\xi_{4i},\xi_{6i}) =\displaystyle\frac{1}{2m}\sum_{i=1}^{2m}\frac{1}{2}\log \left(2\lambda S_w(\theta_i)-2\tilde{\nu}_i\right)+\lambda\mathbb{S}_{w}(\theta_i)-\xi_{4i}-\xi_{6i}-\lambda P+\eta_0+\frac{1}{2},
\end{array}
$$
and $A(\theta)$ and $B(\theta)$ are defined in Theorem \ref{strongdual.thm}.

Further analysis leads to the following consequences. If optimal dual $\lambda, \eta_0, \eta$ lead to $r_{1i}=r_{2i}=0$ for some $i$, in order to maximize $g^D$, we must have the corresponding optimal $\tilde{\nu}_i=0$, and the optimal $\xi_{4i}=\xi_{5i}=\xi_{6i}=0$. Thus $\Xi_i=0$. { The resulting optimal $i$-th component of the cost, denoted by $g_i^D$, is then given by }
$$g_i^D = \frac{1}{2m}\left(\frac{1}{2}\log(2\lambda S_{w}(\theta_i)) +\lambda S_{w}(\theta_i)-\lambda P+\eta_0+\frac{1}{2}\right).
$$
On the other hand, if the optimal $\tilde{\nu}_i$ is $\tilde{\nu}_i>0$ for some $i$ then, using Schur complement on $\Xi_i\geq 0$, the LMI $\Xi_i\geq 0$ is equivalent to
$$
\left[\begin{array}{cc}
\xi_{4i} &\xi_{5i}\cr \xi_{5i} &\xi_{6i}
\end{array}\right]\geq \left[\begin{array}{c}
-\frac{r_{1i}}{2}\cr -\frac{r_{2i}}{2}
\end{array}
\right]\tilde{\nu}_i^{-1}\left[\begin{array}{cc}
-\frac{r_{1i}}{2} & -\frac{r_{2i}}{2}.
\end{array}
\right].
$$
In this case the optimal $\Xi_i$ is rank one and can be obtained by substituting the above inequality into (\ref{xi.eq}). Namely,
\begin{equation}\label{xi2.eq}
\Xi_i=
\left[\begin{array}{c}
\sqrt{\tilde{\nu}_i}\cr
\displaystyle-\frac{r_{1i}}{2\sqrt{\tilde{\nu}_i}}\cr
\displaystyle-\frac{r_{2i}}{2\sqrt{\tilde{\nu}_i}}\cr
\end{array}\right]
\left[\begin{array}{ccc}
\sqrt{\tilde{\nu}_i} &
\displaystyle-\frac{r_{1i}}{2\sqrt{\tilde{\nu}_i}} &
\displaystyle-\frac{r_{2i}}{2\sqrt{\tilde{\nu}_i}}
\end{array}\right]\geq 0.
\end{equation}
The optimal $i$-th component of the cost is then given by the following expression
$$
g_i^D = \frac{1}{2m}\left(\frac{1}{2}\log(2\lambda S_{w}(\theta_i)-2\tilde{\nu}_i) +\lambda S_{w}(\theta_i)-\frac{r^2(\theta_i)}{4\tilde{\nu}_i}-\lambda P+\eta_0+\frac{1}{2}\right),
$$
where $r^2(\theta_i)=r_{1i}^2+r_{2i}^2$.
Therefore, $u_n(m)$ in (\ref{opt_upperbound_approximate}) coincides with the cost of (\ref{cfbhfd5_dual.eq}) and of (\ref{cfbhfd5.eq}) or (\ref{cfbhfd.eq}). Note that $\nu_i$ in $\tilde{g}_m$ of (\ref{opt_upperbound_approximate}) is equal to $2\tilde{\nu}_i$ in $g^D$, while the other dual variables $\eta_0$ and $\eta$ are the same between the two problems. The result then follows from Lemma \ref{lemma_LMI}.

\subsection{Proof of Corollary \ref{Cor}} \label{coro.discrete.dual.val}

From the proof of Lemma \ref{lemma:lower_convergence_equivalent_opt01}, (\ref{xi.eq}) and the complementary slackness condition require that
$$
\left[\begin{array}{ccc} \tilde{\nu}_{i} & -\frac{r_{1i}}{2} & -\frac{r_{2i}}{2}\cr -\frac{r_{1i}}{2} & \xi_{4i} &\xi_{5i}\cr -\frac{r_{2i}}{2} & \xi_{5i} & \xi_{6i}\end{array} \right]\left[\begin{array}{ccc} W_{i} & x_i & b_i\cr x_i & 1 &0\cr b_i & 0 & 1\end{array} \right]={\bf 0}.
$$
When the optimal dual solution is such that $\tilde{\nu}_i>0$, then from (\ref{xi2.eq}) this condition reduces to
$$
\left[\begin{array}{ccc}
\sqrt{\tilde{\nu}_i} &
\displaystyle-\frac{r_{1i}}{2\sqrt{\tilde{\nu}_i}} &
\displaystyle-\frac{r_{2i}}{2\sqrt{\tilde{\nu}_i}}
\end{array}\right]\left[\begin{array}{ccc} W_{i} & x_i & b_i\cr x_i & 1 &0\cr b_i & 0 & 1\end{array} \right]={\bf 0}.
$$
which when solved leads to $W_i=x_i^2+b_i^2$, $a_i=\displaystyle\frac{r_{1i}}{\nu_i}-1$ and $b_i=\displaystyle\frac{r_{2i}}{\nu_i}$, where $\nu_i=2\tilde{\nu}_i$. { Since $c_i = W_i -x_i^2 -b_i^2$ in Corollary \ref{Cor},  $W_i=x_i^2+b_i^2$ indicates $c_i=0$. Namely, when the optimal $\tilde{\nu}_i>0$, then $c_i=0$.} In other words, the optimal primal solution is exclusively obtained by feedback and the feedforward contribution must be zero. Moreover, in this case, the primal optimal solution (\ref{thm_xy_dual_approx}), the components of $\mathbb{Q}$, can be computed from the dual solution. Namely
\begin{equation}
\begin{split}
a_i&=\frac{2\lambda S_{w}(\theta_i)+\eta'A(\theta_i)+\eta_0}{\nu_i}-1, \\
b_i&=\frac{\eta'B(\theta_i)}{\nu_i},\\
\end{split}
\end{equation}
where $\nu_i=2\tilde{\nu}_i$.

\bibliographystyle{IEEEtran}
\bibliography{ref}

\end{document}